\documentclass[useAMS,usenatbib,a4]{mn2e}
\usepackage{epsfig}
\usepackage{times}

\newcommand{\mk}{}

\title[Axisymmetric magnetic fields in stars]{Axisymmetric magnetic fields in stars: relative strengths of poloidal and toroidal components}
\author[Jonathan Braithwaite]{Jonathan Braithwaite \thanks{E-mail: jon@cita.utoronto.ca} \\Canadian Institute for Theoretical Astrophysics\\60 St. George Street, Toronto ON M5S 3H8, Canada}

\pagerange{\pageref{firstpage}--\pageref{lastpage}} 
\begin{document}
\maketitle
\label{firstpage}

\begin{abstract}
In this third paper in a series on stable magnetic equilibria in stars, I look at the stability of axisymmetric field configurations and in particular at the relative strengths of the toroidal and poloidal components. Both toroidal and poloidal fields are unstable on their own, and stability is achieved by adding the two together in some ratio. I use Tayler's (1973) stability conditions for toroidal fields and other analytic tools to predict the range of stable ratios and then check these predictions by running numerical simulations. It is found that while the poloidal field can account for no more than approximately $80\%$ of the total energy, it can account for
a very small fraction of the energy,
i.e. that the toroidal field can be -- and is likely to be -- significantly stronger than the poloidal.
Furthermore, the weaker the field, the weaker the poloidal component can be in relation to the toroidal.
The implications of this result are discussed in various contexts such as the emission of gravitational waves by neutron stars, free precession, and a `hidden' energy source for magnetars.
\end{abstract}

\section{Introduction}
\label{sec:intro}


Magnetic fields are observed in various types of star which are considered unlikely to harbour a suitable regenerative dynamo process because of the lack of convection. For instance, strong fields ($300$ G to $30$ kG) are observed via the Zeeman effect on chemically-peculiar main-sequence A stars (the Ap stars; see \citealt{Mathys:2001} for a review), as well as higher-mass O and B stars. These stars do contain a small convective core and a magnetic field produced inside it could in principle rise through the radiative envelope in the form of buoyant flux tubes, but the time-scale of this process is almost certainly too long for anything to be seen on the surface within the star's lifetime \citep{MacandCas:2003}. The magnetic white dwarfs (WDs, with observed fields of $10^4$ - $10^9$ G) have only weak surface convection, and neutron stars (NSs, with fields $10^8$ - $10^{15}$ G) no convection at all. All of these stars must therefore contain a stable `fossil' magnetic field, either inherited from the original molecular gas cloud or from the previous stage of evolution (e.g. white dwarfs from the main sequence, see \cite{WicandFer:2005}, or neutron stars from a degenerate stellar core, see \cite{FerandWic:2006}) or left over from some kind of dynamo process at the time of formation, either during the pre-main-sequence phase, in the case of Ap stars, or during the convective proto-neutron star phase (\cite{DunandTho:1992}). To have survived since this time, a field must be stable on a dynamic (Alfv\'en) timescale\footnote{The solid crust of a NS can hold an otherwise unstable field in place, provided that the field strength is below some threshold $\sim10^{13}$ G, but it is not clear whether the crust forms soon enough after the end of the proto-NS convective phase to prevent relaxation into stable equilibrium.}. 

It was suggested by Prendergast (1956) that a stellar magnetic field in stable equilibrium must contain both poloidal (meridional) and toroidal (azimuthal) components, since both are unstable on their own. It was then shown more rigorously by Tayler (1973) that any purely toroidal field configuration is unstable at at least some place in a star, and by Wright (1973) and Markey \& Tayler (1973, 1974) that any purely poloidal field is unstable. More recently, the properties of these instabilities of toroidal and poloidal fields have been looked at analytically and numerically (\citealt{Spruit:1998}, \citealt{BraandSpr:2006}, Braithwaite 2006a, b, 2007, \citealt{BonandUrp:2008}).

\begin{figure}
\includegraphics[width=1.0\hsize,angle=0]{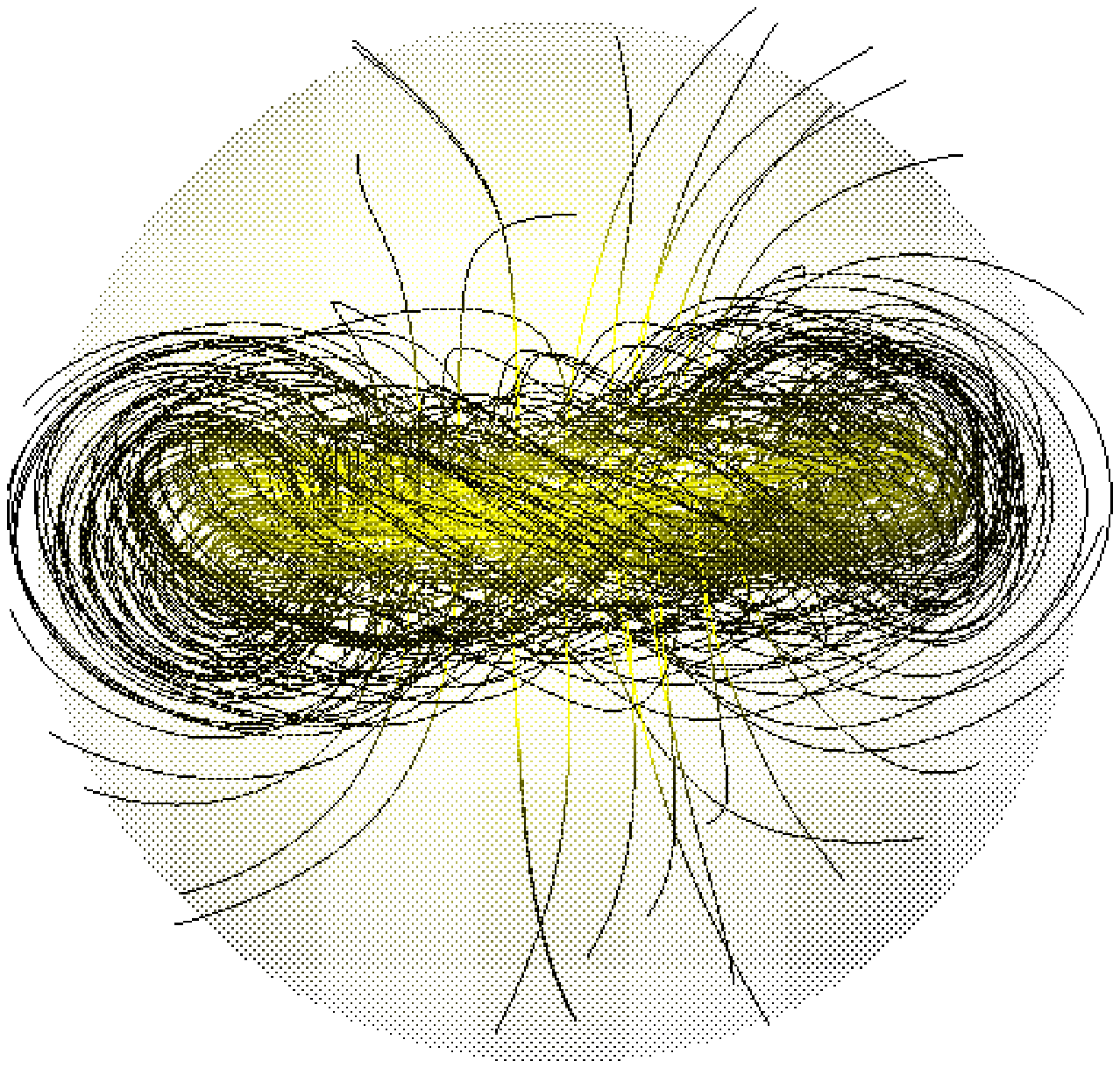}
\includegraphics[width=1.0\hsize,angle=0]{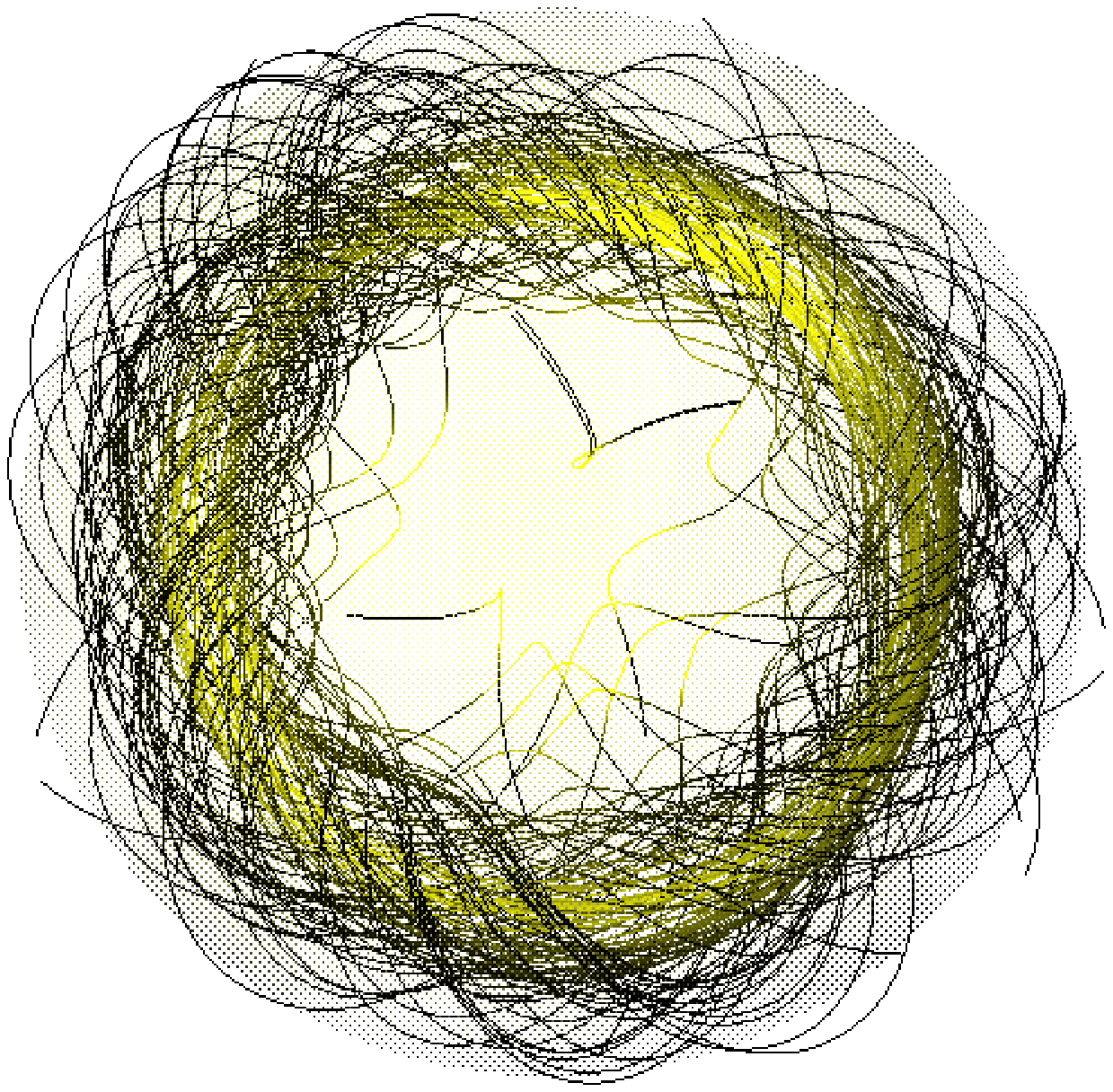}
\caption{The shape of the stable twisted-torus field in a star, viewed from different angles. The transparent surface represents the surface of the star; strong magnetic field is shown with yellow field lines, weak with black.}
\label{fig:torus}
\end{figure}

Analytic methods have proven useful in demonstrating the instability of various equilibrium configurations, but have not been as useful in demonstrating the existence of any stable configuration. Numerical simulations (\citealt{BraandNor:2006}, hereafter Paper I; see also \citealt{BraandSpr:2004}) showed that an arbitrary initial magnetic field inside a non-convective star can evolve on an Alfv\'en timescale into a stable configuration. A roughly axisymmetric configuration was found, of a mixed poloidal-toroidal twisted-torus shape as shown in fig.~\ref {fig:torus}. Once formed it continues to evolve as a result of diffusive processes such as finite conductivity, on a much longer timescale. For instance, the diffusion timescale is $\sim 10^{10}$ years for an Ap star; in the case of a neutron star this timescale is much less certain and is a result of Hall drift and other processes as well as Ohmic dissipation. As the field evolves it moves outwards, passing quasi-statically through a series of stable equilibria until, upon reaching the end of the series of axisymmetric equilibria, it changes to a non-axisymmetric equilibrium. These non-axisymmetric equilibria are described in more detail in \citealt{Braithwaite:2008} (hereafter Paper II), where it was also found that a non-axisymmetric equilibrium can be formed on an Alfv\'en timescale directly, from somewhat different initial conditions. Essentially, the important difference is the central concentration of the initial field and the fraction of flux connected through the stellar surface -- a non-axisymmetric equilibrium can be formed directly from an initial field whose energy is more `spread out' rather than concentrated towards the centre of the star, and which has significant flux connection through the stellar surface.


We are concerned here with only the axisymmetric class of equilibria. Since both toroidal and poloidal fields are unstable on their own, there is presumably some allowed range of ratios of the two respective field strengths; it is the principal purposes of this paper shed some light on what these stable ratios might be. The toroidal field is always confined to the interior of a star, since a toroidal field on or above the surface would require long-lived currents outside the star, so that we observe on the surface only the poloidal component. Therefore, it is either difficult or impossible to get any {\it direct} observational constraint on the range of allowed ratios. There are however some interesting ways in which a toroidal magnetic field can manifest itself indirectly, which make the question of allowed poloidal/toroidal ratios a useful line of study. Firstly, in predominantly non-convective main-sequence stars (i.e. $> 1.5 M_\odot$) it would be useful to know how much flux may be hidden below the surface, since this flux may be important during formation, eventually be visible on the WD, be responsible for shaping the planetary nebula, affect the supernova explosion in some way, and affect the natal rotation periods of NSs and WDs via core-envelope coupling. In fact, this reminds us of another question: how much of the {\it poloidal} flux can be hidden below the surface? Certainly not all of it has to go through the stellar surface, but may be confined to the interior. This paper also sheds some light on this issue. In the context of the `magnetars', highly magnetised NSs (dipole field strength on surface $10^{14 - 15}$ G; see \citealt{WooandTho:2004} for a review), it would certainly be useful to know how much magnetic flux and energy could be `hidden' in the interior of the star. These stars undergo soft-$\gamma$-ray outbursts, releasing as much as $~2\times10^{46}$ erg of magnetic energy in less than a second. A field of $3\times 10^{14}$ G, if it fills the interior of the star, contains $~2\times 10^{46}$ erg, and since these stars appear to have a lifetime of $\sim 10^4$ years and to undergo large outbursts perhaps once a century, it seems likely that the average field strength in the interior is significantly greater than that on the surface.

Another way in which the poloidal/toroidal ratio may manifest itself is through its effect on the star's shape and moment of inertia. It has been known for some time (e.g. \citealt{ChaandFer:1953}, \citealt{Wentzel:1961}) that a poloidal field will make a star oblate and a toroidal field prolate, and obviously with a mixed poloidal-toroidal field it will depend on the ratio of the two. In general, such a deformed star should undergo torque-free precession\footnote{Purists may prefer the term `nutation', although `precession' occurs more frequently in the literature.}; there is already some observational evidence for this (\citealt{Cutleretal:2003}, \citealt{Wasserman:2003} and \citealt{Akguenetal:2006}). If this precession is damped, then kinetic energy is minimised while conserving angular momentum and a prolate star will tend towards alignment of its rotation and magnetic axes, while in an oblate star the angle between the two axes will tend to $90^\circ$. In an Ap star, this damping process may or may not take on the order of a main-sequence lifetime \citep{Mesteletal:1981} but in a NS it may be much faster and a predominantly toroidal field in a fast-rotating NS could lead to strong emission of gravitational radiation. These effects of the magnetic field on a star's moment of inertia will be looked at in more detail in a forthcoming paper (Braithwaite \& Nissanke, in prep.)

In section~\ref{sec:analytic}, the instability in toroidal fields is described and some predictions are made about the stability of mixed poloidal-toroidal fields, and the properties of instability in poloidal fields is reviewed. In section~\ref{sec:numerical}, analytic conditions are used to examine the stability of fields produced in simulations, and simulations are presented of the decay or otherwise of fields with various toroidal-poloidal ratios. I conclude, and discuss the results and their applications in section~\ref{sec:discussion}.

\section{Instability in axisymmetric fields: analytic results}
\label{sec:analytic}

In this section I review the nature and properties of instability in purely toroidal and purely toroidal magnetic fields in stars, as well as look at stabilisation of a predominantly toroidal field through the addition of a weak poloidal component and vice versa.

\subsection{The Tayler instability}
\label{sec:analyticTayler}

The stability of purely toroidal fields in stars was examined by \citet{Tayler:1973}, who used the energy method of \citet{Bernsteinetal:1958} to derive necessary and sufficient stability conditions. Given a perturbation of the form
\begin{equation}
\xi \sim f(\theta) e^{i(nr+m\phi)+\sigma t},
\end{equation}
in spherical polar coordinates ($r, \phi, \theta$) he found that the stability conditions for $m\ge2$ are less strict than for $m=1$, so that if the goal is simply to distinguish between unstable and stable field configurations, these higher azimuthal wavenumbers need not be considered. The unstable modes can therefore be thought of as local in the meridional plane but global in the azimuthal direction.

\subsubsection{Necessary and sufficient stability conditions}

For stability against the $m=0$ mode we need the change in potential energy, as given by the following integral, to be positive for an arbitrary displacement field ${\mathbf \xi}$. In cylindrical coordinates $(\varpi, \phi, z)$:

\begin{equation}
\delta W \propto \int \varpi d\varpi dz \{ (......)^2 + a \xi_z^2 + b \xi_z \xi_\varpi + c \xi_\varpi^2 \},
\end{equation}
where $(......)^2$ is some function of ${\mathbf \xi}$ which is positive definite, and where
\begin{equation}
a = g_z \frac{\partial \rho}{\partial z} - \frac{\rho^2 g_z^2}{B_\phi^2+\gamma P},
\end{equation}
\begin{equation}
b = g_\varpi \frac{\partial \rho}{\partial z} + g_z \frac{\partial \rho}{\partial \varpi}
 - \frac{2\rho g_z (\rho g_\varpi - 2 B_\phi^2/\varpi)}{B_\phi^2+\gamma P}
 - \frac{2B_\phi}{\varpi}\frac{\partial B_\phi}{\partial z},
\end{equation}
\begin{equation}
c = g_\varpi \frac{\partial \rho}{\partial \varpi} - \frac{(\rho g_\varpi - 2 B_\phi^2/\varpi)^2}{B_\phi^2+\gamma P}
-\frac{2B_\phi}{\varpi}\frac{\partial B_\phi}{\partial \varpi} + \frac{2B_\phi^2}{\varpi^2}.
\end{equation}
For $\delta W$ to be positive, clearly it is sufficient (and can also be shown to be necessary) that the quadratic form is positive everywhere in the volume of integration. So the stability conditions are that
\begin{equation}
a > 0 \:,\;\;\; c > 0 \;\;\;\;{\rm and} \;\;\;\; b^2 < 4ac
\label{eq:tayler-m0}
\end{equation}
everywhere in the $(\varpi,z)$ plane in the star.

For stability against the $m=1$ mode, we have a similar integrand consisting of one positive term and a quadratic where
\begin{equation}
a = g_\varpi \frac{\partial \rho}{\partial \varpi} - \frac{\rho^2 g_\varpi^2}{\gamma P}
- \frac{B_\phi^2}{\varpi^2} - \frac{2B_\phi}{\varpi}\frac{\partial B_\phi}{\partial \varpi},
\end{equation}
\begin{equation}
b = g_\varpi \frac{\partial \rho}{\partial z} + g_z \frac{\partial \rho}{\partial \varpi}
 - \frac{2\rho^2 g_\varpi g_z}{B_\phi^2+\gamma P}
 - \frac{2B_\phi}{\varpi}\frac{\partial B_\phi}{\partial z},
\end{equation}
\begin{equation}
c = g_z \frac{\partial \rho}{\partial z} - \frac{\rho^2 g_z^2}{\gamma P} + \frac{B_\phi^2}{\varpi^2}.
\end{equation}
Tayler went on to show that it is impossible to satisfy all six of these conditions everywhere in the star, concluding that a stable equilibrium cannot be purely toroidal.

\begin{figure}
\includegraphics[width=1.0\hsize,angle=0]{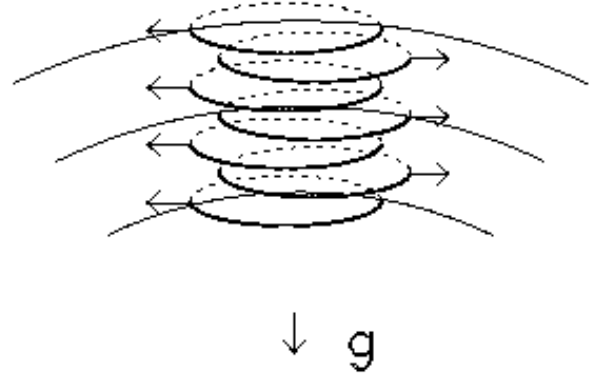}
\caption{The form of the instability in a toroidal field near the axis of symmetry of a star. The $m=1$ mode sets in before $m=0$ and $m\geq2$. [Figure from \citealt{Spruit:1999}.]}
\label{fig:cartoon}
\end{figure}

A hand-waving explanation of the instability mechanism is as follows. One can imagine the toroidal field as a stack of field rings which exert pressure on one another. Magnetic tension (`hoop stress') and external pressure prevents the discs from simply expanding outwards and thus relieving this pressure, but it is energetically favourable for the rings to slip sideways out of the equilibrium position, rather like the way an overloaded spinal column can result in a `slipped disc'. The result is an $m=1$ `kink' mode, as illustrated in fig.~\ref{fig:cartoon}.

\subsubsection{Unstable wavelengths and the stabilising effect of an added poloidal field}
\label{sec:torwithaxial}

The growth rate of the instability can be shown to be (\citealt{Tayler:1957})
\begin{equation}
\sigma \sim \omega_{\rm A} \equiv \frac{v_{\rm A}}{r \sin \theta} = \frac{B_\phi}{r \sin \theta \sqrt \rho}.
\label{eq:growthrate}
\end{equation}
In the case of an unstratified medium with infinite conductivity and zero kinetic viscosity, all vertical wavelengths (i.e. all $n$) are unstable. However, finite conductivity damps the highest $n$ modes faster than they can grow, resulting in a maximum unstable $n$. Conversely, stable stratification restricts the growth of the lowest wavenumbers. This is because, as we see from the continuity equation, the instability results in vertical motions of order $\xi_r \sim \xi_{\rm h} l_r/l_{\rm h}$, where $\xi_r$ and $\xi_{\rm h}$ are the displacements in the vertical (radial) and horizontal directions respectively, and $l_r$ and $l_{\rm h}$ are the length scales in the vertical and horizontal, so that $l_r = 1/n$. The horizontal instability force per unit mass pushing a fluid element away from its equilibrium position is $F_{\rm h} = \sigma^2 \xi_{\rm h}$, and the vertical buoyancy force pushing the fluid element back to equilibrium is $F_r = N^2 \xi_r$, where $N$ is the buoyancy frequency. If the instability is to grow, we need therefore to have $\xi_{\rm h} F_{\rm h} > \xi_r F_r$. Now, combining these two limits we have the following range of unstable wavelengths \citep{Spruit:1999}
\begin{equation}
\sqrt{\frac{\sigma}{\eta}} > n > \frac{N}{l_{\rm h} \sigma}
\label{eq:diffandgrav}
\end{equation} 
where $\eta$ is the magnetic diffusivity. A toroidal field is stablised if the left hand side is less than the right hand side. These wavelength limits were confirmed numerically by \citet{Braithwaite:2006a}.

Adding a radial field component $B_r$ will also help to stabilise a toroidal field, as the horizontal displacements produced by the instability will bend the radial field lines, producing a restoring force. Using similar arguments to those in the previous paragraph, we can predict what field strength will be required for stabilisation. The (horizontal) restoring force per unit mass is $\xi_{\rm h} B_r^2 / ( l_r^2 \rho )$ and, equating this to the force from the instability, we have instability if
\begin{equation}
\frac{\sigma \sqrt{\rho}}{B_r} \sim
\frac{B_\phi}{r \sin\theta B_r} > n,
\label{eq:bz}
\end{equation}
using (\ref{eq:growthrate}). This corresponds to a result in the case of an unstratified plasma from the unstable-modes and dispersion-relation method \citep{Tayler:1957}.

Taking numbers typical for a main-sequence star (in cgs units) $r\sin\theta\sim10^{11}$, $\rho\sim1$, $B_r\sim B_\phi\sim10^3$ and $\eta\sim10^2$, the term on the left of (\ref{eq:diffandgrav}) turns out to be $10^6$ times greater than that in ({\ref{eq:bz}). Similarly high ratios are found in WDs and NSs. This means that we can expect the stabilising effect of the poloidal field to dominate over that of magnetic diffusion. Therefore, ignoring magnetic diffusion and combining the effects of stable stratification and poloidal field, we have stability if
\begin{equation}
\sigma^2 - \frac{N^2}{n^2 l_{\rm h}^2} - \frac{B_r^2 n^2}{\rho} < 0,
\label{eq:stablelambda}
\end{equation}
which is true for all wavenumbers $n$ if 
\begin{equation}
\frac{B_\phi^4}{\varpi^4} < \frac{4B_r^2 \rho N^2}{l_{\rm h}^2},
\label{eq:stabcond}
\end{equation}
which can be rewritten in terms of the Alfv\'en frequency $\omega_{\rm A} \equiv B_\phi/(\varpi\sqrt{\rho})$ as
\begin{equation}
\frac{\omega_{\rm A}}{N} \cdot \frac{l_{\rm h}}{2\varpi} < \frac{B_r}{B_\phi}.
\label{eq:stabcond2}
\end{equation}
The second part of the left hand side will be roughly unity, so it is clear that the critical $B_r/B_\phi$ depends on the field strength in the star. In a main-sequence star, we have $\omega_{\rm A}^2/N^2 \sim E/U$, the ratio of magnetic to thermal energies; in neutron stars the buoyancy is weaker \citep{Reisenegger:2008}, so that $\omega_{\rm A}^2/N^2 \sim 10^2 E/U$. However, we expect $E/U < 10^{-6}$ in a real star, so a relatively weak poloidal field should be sufficient for stabilisation in any kind of star.

\subsection{Instability of poloidal fields}

It is known that not only purely toroidal fields, but also purely poloidal fields are also unstable (\citealt{MarandTay:1973}, \citealt{Braithwaite:2007}). The instability begins in the region of the neutral line, the line where the poloidal component goes to zero. This instability is not unlike that in a toroidal field near the axis of symmetry -- the poloidal field increases in proportion to distance from the neutral line and it is the pressure in the direction parallel to this line, which the `loops' are exerting on each other, which drives the instability. The direction of the stratification is the important difference between the two instabilities; displacements must be approximately perpendicular to gravity, which in the poloidal case means that the loops move in a direction parallel to the star's axis of symmetry. This is illustrated in fig.~\ref{fig:pol_inst}. The growth rate of this instability is calculated in the same way as in the toroidal case, so it is given by the local Alfv\'en frequency around the neutral line $\omega_{\rm A} \equiv B_{\rm p}/(s\sqrt{\rho})$ where $s$ is the distance from the neutral line. Adding a toroidal component (i.e. a component parallel to the neutral line) can stabilise the field, since the instability increases the length of the neutral line, thereby stretching the toroidal field line lying on it. The higher azimuthal wavenumbers are stabilised first, so that the $m=2$ wavenumber is the first to become unstable as the strength of the toroidal component is reduced. [Note that the $m=0$ and $m=1$ modes are prevented by conservation of momentum and angular momentum respectively.] In Paper II it was found that as an axisymmetric configuration diffuses outwards towards the stellar surface, the toroidal component is lost into the atmosphere, weakening it relative to the poloidal component, and eventually the $m=2$ mode appears and the configuration evolves quasi-statically along a non-axisymmetric sequence. The minimum-energy state of this non-axisymmetric equilibrium has comparable toroidal and poloidal components, which strongly suggests that an axisymmetric equilibrium cannot have a much stronger poloidal component than toroidal. A poloidal component which is stronger by some factor of order unity is possible, and the next section examines how large that factor can be. Note that this signals an asymmetry between the toroidal and poloidal cases, as there is no equivalent factor-of-order-unity argument giving an upper bound to the relative strength of the toroidal component.

\begin{figure}
\includegraphics[width=1.0\hsize,angle=0]{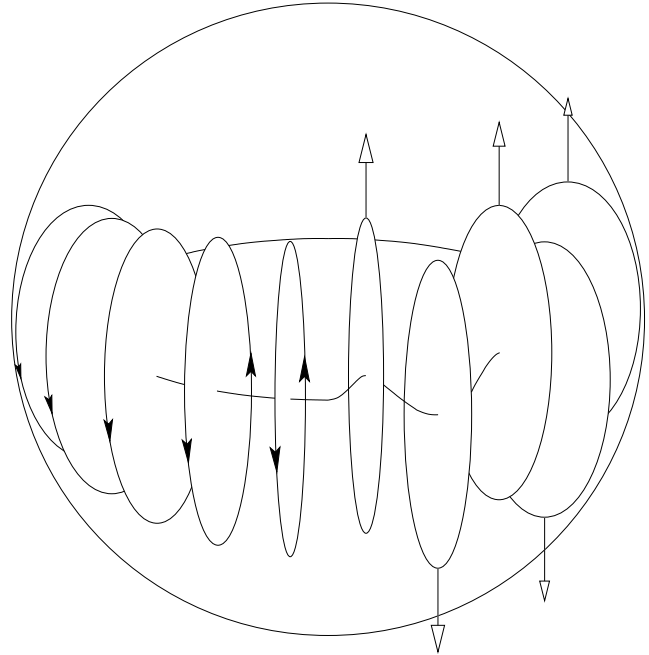}
\caption{The form of the instability in a purely poloidal configuration. The left side shows the equilibrium and the right side the growth of a mode of particular azimuthal wavenumber.}
\label{fig:pol_inst}
\end{figure}

The strength of toroidal field required for stabilisation can be estimated with an analogous method to that described in section \ref{sec:torwithaxial}, taking the wavelength to be $2\pi r_{\rm n}/m$ and arriving at the following stability condition:
\begin{equation}
\frac{B_{\rm p}}{s B_{\rm t}} < \frac{m}{r_{\rm n}},
\label{eq:polfrac}
\end{equation}
where $s$ is the distance from the neutral line and $r_{\rm m}$ is the radius of the neutral line. Note that $B_{\rm p}$ is proportional to $s$ in the neighbourhood of the neutral line. Although this stability condition should be taken as rather approximate since curvature and other effects have been ignored, it shows again that stability against the (most unstable) $m=2$ mode requires the toroidal component to be at least comparable in strength to the poloidal component. \citet{Wright:1973} arrived at a similar stability condition (his eqn.~39):
\begin{equation}
0.24 B_{\rm p} (s_0) < B_{\rm t}
\label{eq:wright}
\end{equation}
where $s_0$ is some distance from the neutral line whose value is uncertain. Note that unlike in the case of strong toroidal fields in the previous section, the critical ratio here of toroidal to poloidal field strengths {\it does not depend} on the absolute magnetic energy in relation to the thermal energy and buoyancy properties of the star.

\subsection{The effect of rotation}\label{rotation}

{\mk In the previous sections the rotation of the star was not taken into consideration, but we may now look at its possible effect. First of all, we expect it to have an effect only in the `fast' rotating regime, i.e. when the rotation speed is faster than the Alfv\'en speed \citep{FriandRot:1960}. Applying this condition to real astrophysical objects, we see some fast rotators and some slow rotators. We find the ratio of the magnetic and rotation timescales to be
\begin{equation}
\frac{\Omega}{\omega_{\rm A}} = \frac{2\pi}{P\bar{B}}\sqrt{\frac{3M}{R}}
\end{equation}
where $P$ is the rotation period, $\bar{B}$ is some average magnetic field in the interior of the star, and $M$ and $R$ are the stellar mass and radius. For $3M_\odot$ MS stars, $0.8M_\odot$ WDs and $1.4M_\odot$ NSs, the numbers are
\begin{equation}
\frac{\Omega}{\omega_{\rm A}} \approx
3800 P^{-1}_{10{\rm d}} B^{-1}_{5{\rm kG}} \,\,\,\,{\rm or}\,\,\,\,
18 P^{-1}_{1{\rm d}} B^{-1}_{7} \,\,\,\,{\rm or}\,\,\,\,
0.05P^{-1}_{10{\rm s}} B^{-1}_{15}
\end{equation}
where the periods are in units of $10$ days, $1$ day, $10$ seconds and the field strengths $5$kG, $10$MG and $10^{15}$G -- typical values for magnetic MS stars, WDs and NSs (magnetars) respectively. Evidently the typical Ap star is in the fast rotating regime; however there are a few known examples with comparable field strengths but rotation periods of $30-100$ years and above, just putting them into the slowly rotating regime \citep{Mathys:2008}. There is also much variation amongst the magnetic WDs: a group exists with $P\sim 100$yr and $B\sim 300$MG while some others have $P\sim$ few hours and $B\sim 1$MG \citep{FerandWic:2005}, so this class of star contains both fast and (very) slow rotators. The magnetars show much less variation and all are slow rotators; the radio pulsars on the other hand are fast rotators.

In the fast rotating stars, the rotation will certainly have some effect on the growth rate of any instability. In the rotating frame, the Coriolis force acts perpendicularly to any unstable displacement, giving rise to epicycles; in the absence of any diffusive mechanisms this causes the fluid elements to come back to their original positions and stabilisation results. This has been seen in simulations of the instability in a toroidal field \citep{Braithwaite:2006a}. Diffusion though damps the epicycles and the instability returns, albeit with the growth rate reduced by some factor $\Omega/\omega_{\rm A}$ (\citealt{PitandTay:1986}, \citealt{Spruit:1999}). In the case of a strong poloidal field, a similar effect has been seen in simulations \citep{Braithwaite:2007}. It seems likely then that rotation does not affect the upper and lower bounds on $E_{\rm p}/E$.
}


\section{Stability of axisymmetric field configurations: numerical methods}
\label{sec:numerical}

Having
derived 
some stability criteria for a mixed poloidal-toroidal field, it remains to see what this means for the stability of a global magnetic field configuration. Clearly, a globally stable configuration needs to be locally stable against the Tayler instability at each point in the star, as well as being stable in the neighbourhood of the neutral line. There are two obvious ways to proceed. The first is to find a global configuration and then check that a local analysis predicts stability at every location in the star for both types of instability. The second way is to construct a global configuration and then numerically follow its evolution in time.

\subsection{Constructing an axisymmetric field}

The basis for constructing axisymmetric field configurations whose stability we can probe will be the result of numerical simulations similar to those performed in Paper I, to where the reader is referred for a fuller account of the setup of the model; a brief outline is given here.

The code used is the {\sc stagger code} (\citealt{NorandGal:1995}, \citealt{GudandNor:2005}), a high-order finite-difference Cartesian MHD code which uses a `hyper-diffusion' scheme, a system whereby diffusivities are scaled with the length scales present so that badly resolved structure near the Nyquist spatial frequency is damped whilst preserving well-resolved structure on longer length scales. This, and the high-order spatial interpolation and derivatives (sixth order) and time-stepping (third order) increase efficiency by giving a low effective diffusivity at modest resolution ($128^3$ here). The code includes Ohmic and well as thermal and kinetic diffusion. Using Cartesian coordinates avoids problems with singularities and simplifies the boundary conditions: periodic boundaries are used here.

The simulations model the star as a self-gravitating ball of ideal gas ($\gamma=5/3$) of radius $R$ in hydrostatic equilibrium with radial density and pressure profiles obeying the polytrope relation $P \propto \rho^{1+\frac{1}{n}}$, with the index $n$ set to $3$ here, since this gives stable stratification and is a reasonable approximation to an upper-main-sequence star. It seems unlikely that a different EOS, for instance that of a neutron star, will make even much quantitative difference to the results. The important point is the stable stratification -- the issue of magnetic equilibria in a non-stably-stratified star will be explored in a forthcoming paper.

{\mk A boundary between the star and the surrounding volume is produced in the following way. In reality, we expect the field outside the star to be not only force-free (as is the usual approximation when looking at magnetic loops in the solar corona) but also curl-free, since a very tenuous medium will not sustain long-lived currents. Numerically, a potential (i.e. current-free) field is tricky to produce, but fortunately the same effect can be produced by adding a high magnetic diffusivity to the volume outside the star, and in practice the diffusivity chosen is the maximum allowed without having to reduce the timestep. This causes the field outside the star to relax fairly rapidly to a curl-free configuration. Also, the gas outside the star is hot, increasing the scale height and thus stopping the density from falling to greatly towards the edges of the computational box, preventing high Alfv\'en speeds and the lower timestep they would cause.}
The star is given an initially random magnetic field containing energy at all length scales down to a limit of a few grid-spacings, and the MHD equations are integrated in time to follow the evolution of the field. Within a few Alfv\'en crossing-times, a stable equilibrium is reached. In the case where the field is less centrally concentrated, the equilibrium is non-axisymmetric, consisting of twisted flux tubes meandering at roughly constant depth under the surface of the star. This case was examined in detail in Paper II. In the case of more centrally-concentrated (i.e. deeply-buried) initial fields, an approximately axisymmetric field forms. Ignoring the small deviations from axisymmetry, there appear to be two basic degrees of freedom\footnote{There are other, more subtle degrees of freedom concerning the exact shape of the field lines, but they appear to be less important, and are beyond the scope of this study.}: the concentration of the field into the centre of the star, which can be parametrized as $r_{\rm n}$, the distance between the axis of symmetry and the neutral line, and the poloidal fraction of total energy $E_{\rm p}/E$. In this two-dimensional parameter space lies an area of stability through which the star slowly moves in time, as a result of diffusive processes such as finite conductivity, to ever increasing $r_{\rm n}$
and the eventual end of its stable axisymmetric life, as described above and in Paper II. It is the aim here to find the boundaries of that area of stability.

The first step is to produce a stable field in a simulation (i.e. run with arbitrary initial conditions for a few Alfv\'en crossing times until the field has settled down into an equilibrium) and axisymmetrise it, using an axis defined by $\int \mathbf{r} \times \mathbf{B} dV$. Although the fields produced in the simulations are already approximately axisymmetric, perfect axisymmetry simplifies the stability analysis. Also, the small asymmetry between the two hemispheres is removed so that the field is symmetrical about the $z=0$ plane. Now, this symmetrised star can be put back into the simulation and evolved in time (the `fiducial simulation'), where it slowly diffuses outwards, the value of $r_{\rm n}$ rising as it does so. Its stability can then be examined at various points along this diffusive evolution track, both analytically, using the methods described in Sect.~\ref{sec:analytic}, and numerically, by changing the relative strengths of the poloidal and toroidal parts and using that as the initial conditions for a new simulation. In this way, the boundaries of the stable area in $r_{\rm n}$ -- $E_{\rm p}/E$ parameter space can be found. [In the simulations described here, the ratio of magnetic to thermal energies in the star $E/U = 1/400$ unless stated otherwise.] In fig.~\ref{fig:axisym} we see the axisymmetric field at the times $t/\tau_{\rm A}=5.2$, $69.7$ and $97.0$ when $r_{\rm n}/R=0.33$, $0.47$ and $0.58$. Time is measured in units of the Alfv\'en crossing time, defined here as $\tau_{\rm A}\equiv R\sqrt{M/2E}$ where $E$ is the total magnetic energy and $M$ is the mass of the star. Note that it can be seen in the figure that the contours of $\varpi B_\phi$ are parallel to the poloidal field lines. This result can be derived analytically (\citealt{Mestel:1961}, \citealt{Roxburgh:1966}) from $\nabla\cdot{\bf B}=0$ and from the recognition that the azimuthal component of the Lorentz force must be zero everywhere because in an axisymmetric equilibrium it cannot be balanced by gravity $\rho {\mathbf g}$ or the pressure gradient $-\nabla P$.

\begin{figure}
\includegraphics[width=0.32\hsize,angle=0]{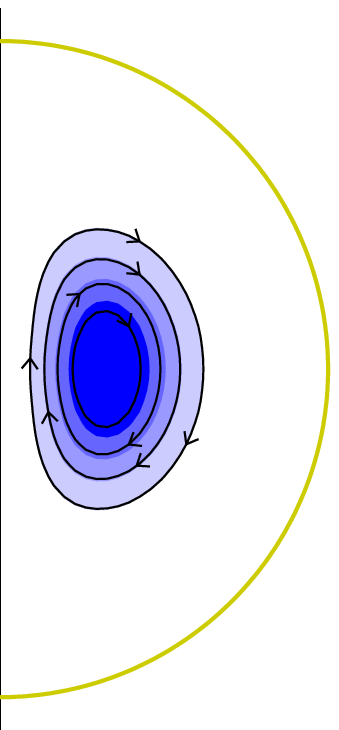}
\includegraphics[width=0.32\hsize,angle=0]{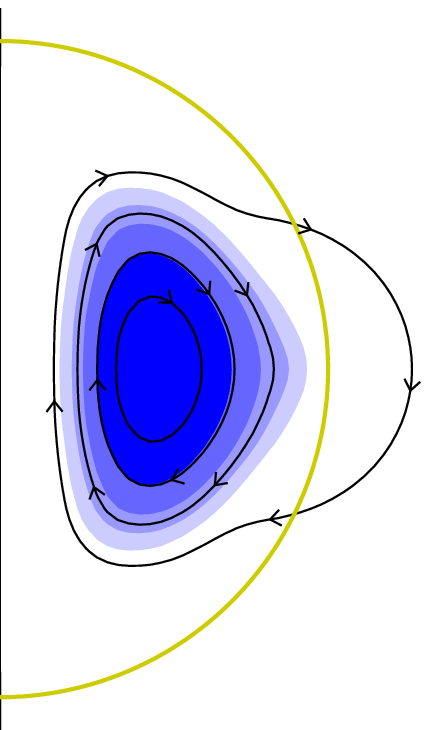}
\includegraphics[width=0.32\hsize,angle=0]{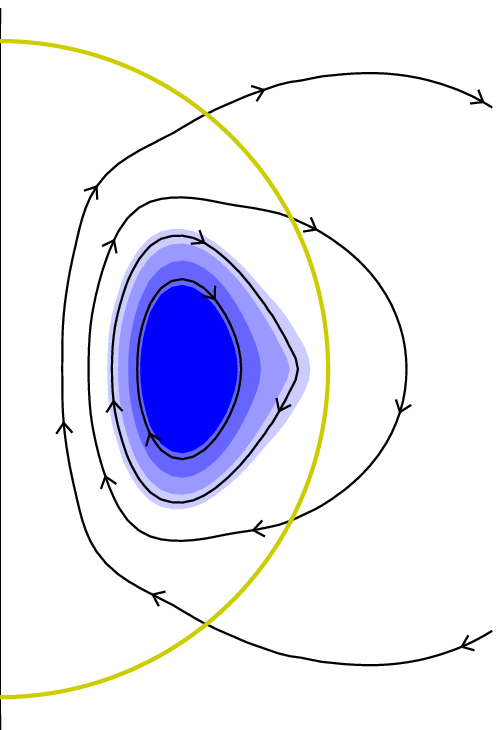}
\caption{Projection onto the meridional plane of stable magnetic field configurations with $r_{\rm n}/R=0.33$, $0.47$ and $0.58$. The yellow semicircle is the surface of the star, and the black lines and blue shading represent the poloidal and toroidal components of the field, respectively. The contours of the toroidal part are actually contours of $\varpi B_\phi$; it is clear that this quantity is roughly constant on poloidal field lines. The poloidal lines plotted are separated by equal quantities of poloidal flux.}
\label{fig:axisym}
\end{figure}

In fig.~\ref{fig:all_vs_t} are plotted various quantities from the fiducial run which change in time as the magnetic field evolves diffusively: the neutral line radius $r_{\rm n}$, the poloidal energy fraction $E_{\rm p}/E$ and the fraction of the poloidal flux which breaches the surface of the star $\Phi_{\rm surf}/\Phi_{\rm p}$.

\begin{figure}
\includegraphics[width=1.0\hsize,angle=0]{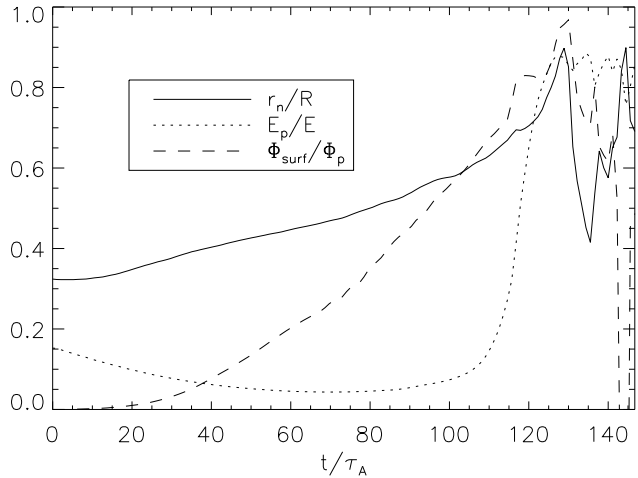}
\caption{The neutral line radius $r_{\rm n}$, the poloidal energy fraction $E_{\rm p}/E$ and the fraction of poloidal flux which breaches the surface of the star $\Phi_{\rm surf}/\Phi_{\rm p}$ against time (in units of the Alfv\'en crossing time). Transition to non-axisymmetric equilibrium occurs at around $t/\tau_{\rm A}=120$.}
\label{fig:all_vs_t}
\end{figure}

First, the case where the toroidal field dominates will be examined, and then in section~\ref{sec:poldom} the poloidal-dominated case is looked at, in order to find both boundaries of stability, i.e. both upper and lower bounds to $E_{\rm p}/E$.

\subsection{Stability of a predominantly toroidal field}

In this section I examine the stability of fields with low $E_{\rm p}/E$, firstly with a combination of Tayler's stability conditions and equation (\ref{eq:stabcond}), and secondly with simulations using different values of $E_{\rm p}/E$ as initial conditions.

\subsubsection{Analytic stability conditions}\label{sec:tortayler}

The $m\geq2$ modes always set in after the $m=1$ mode and need not be considered here. Taking the output the fiducial simulation (where $E/U = 1/400$), described in the previous section, I apply Tayler's six conditions (\ref{eq:tayler-m0}) for both $m=0$ and $m=1$ modes to the toroidal component of the field, at the three points in time shown in fig.~\ref{fig:axisym}. Looking at where in the star the conditions are met, the first result is that the first two conditions, $a>0$ and $c>0$ for both the $m=0$ and $m=1$ modes are satisfied everywhere in the star at all points in time. In contrast, the two $b^2 < 4ac$ conditions are not met everywhere; the regions in the meridional plane where they are and are not satisfied this are plotted in fig.~\ref{fig:stabcond}, for the case where $r_{\rm n}=0.47R$. Of course, this is the conclusion that Tayler arrived at -- that a purely toroidal field is always unstable in at least some part of the star. The field which forms in the simulations is stable against the Tayler instability only because of the presence of the poloidal component. However, it is possible that the toroidal component can be stabilised with a weaker poloidal component than is actually present in the fiducial simulation. We can use (\ref{eq:stabcond}) to estimate what strength the poloidal field needs to be at any particular point, so an estimate for the minimum overall strength of the poloidal component is that at which this condition is just satisfied at every point where Tayler's criteria are not satisfied. It should be stressed however that this approach is only approximate, since we have ignored curvature effects, and also because $l_{\rm h}$ is undetermined -- we shall simply assume here that $l_{\rm h} \sim R$. The area stabilised according to (\ref{eq:stabcond}) is also plotted in fig.~\ref{fig:stabcond}. It is clear from the figure that the entire region unstable to $m=1$ is stabilised by the poloidal field, as well as almost the entire region otherwise unstable to $m=0$, but that two small regions are apparently still $m=0$ unstable. However, looking at the minimum unstable wavelength given by (\ref{eq:stablelambda}), we see that both of these areas, there is barely space to fit one wavelength as $\lambda_{\rm min}\sim2dx\approx0.06R$. This explains the stability in the simulations at this poloidal/toroidal ratio. Note that the reason for expressing the minimum and maximum unstable wavelengths in terms of the grid spacing is that we can easily check if the simulation is running at sufficiently high resolution -- clearly the wavelength needs to be at least a few grid spacings to be properly resolved.

\begin{figure}
\includegraphics[width=1.0\hsize,angle=0]{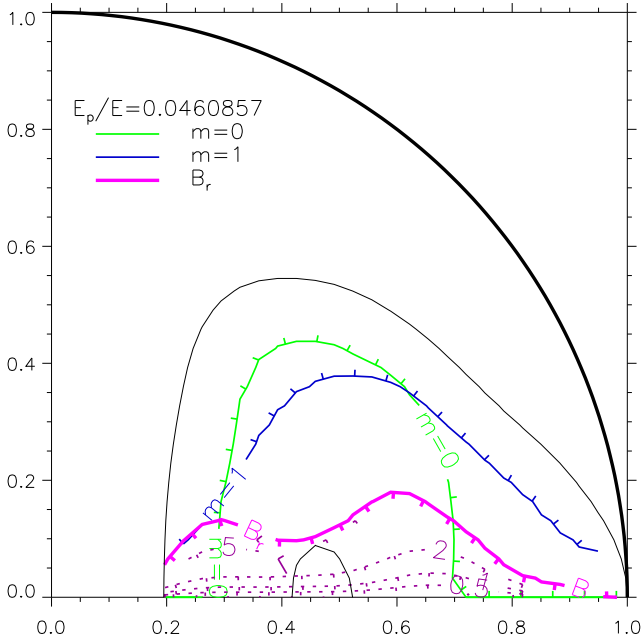}
\caption{Half of the meridional plane (the other half being identical) with $r_{\rm n}/R=0.47$ value. The thick black line is the surface of the star and the thin black lines are two selected poloidal field lines, showing the locations of the `closed' poloidal region (where the toroidal field resides) and of the neutral line. The regions stable against the $m=0$ and $m=1$ even in the absence of a poloidal field are illustrated by the green and blue lines; the ticks point into the region of instability. The thick pink line shows which area is stabilised by the radial component of the field $B_r$. Dotted and dashed purple lines are contours of minimum and maximum unstable wavelengths, in units of the grid spacing $dx\approx0.03R$, as calculated by solving for $n$ in (\ref{eq:stablelambda}) and taking $\lambda=2\pi/n$.}
\label{fig:stabcond}
\end{figure}

If the relative strengths of the toroidal and poloidal components are now changed (while keeping the total energy $E$ constant), the position of the $B_r$ stabilisation line changes\footnote{In principle the $m=0$ and $m=1$ stability lines also move, but in the case of strong toroidal field and constant total energy, the toroidal field strength changes only by a small factor and the lines move very little.}. Fig.~\ref{fig:gall.fracs} shows the positions of the stability lines at various ratios $E_{\rm p}/E$ of poloidal/total energy at the three different values of $r_{\rm n}$.

Looking first at the top row of fig.~\ref{fig:gall.fracs} ($r_{\rm n}/R=0.33$, $E_{\rm p}/E=0.139$ in the fiducial simulation), it can be seen that the $E_{\rm p}/E=0.01$ should be unstable, as there is an $m=1$ unstable region on the left where the minimum unstable wavelength is only a few grid spacings, small enough to fit into the space available (but large enough to be resolved numerically in the simulations described below). However, the minimum wavelength looks too high in the $m=0$ unstable zone. The $0.018$ case is more marginal but there may still be space to fit an unstable wavelength of 5 grid spacings ($~0.15R_\ast$) into the $m=1$ unstable space. The $0.032$ case looks also marginal, but $0.056$ does look stable. Therefore, the critical value of $E_{\rm p}/E$ is perhaps around $0.018$ or $0.032$. Repeating this analysis on the middle row of fig.~\ref{fig:gall.fracs} ($r_{\rm n}/R=0.47$, $E_{\rm p}/E=0.046$ in the fiducial simulation) we get a stable $E_{\rm p}/E$ minimum at around $E_{\rm p}/E=0.0056$ to $0.01$. Finally on the bottom row of fig.~\ref{fig:gall.fracs} ($r_{\rm n}/R=0.58$, $E_{\rm p}/E=0.066$ in the fiducial simulation) we see that practically the entire region of interest is $m=0$ unstable, which is stabilised by the poloidal field at around $E_{\rm p}/E=0.056$, but that the $m=1$ mode should be stabilised at a somewhat lower poloidal energy ratio, perhaps at roughly $0.01$.

\begin{figure*}
\includegraphics[width=0.246\hsize,angle=0]{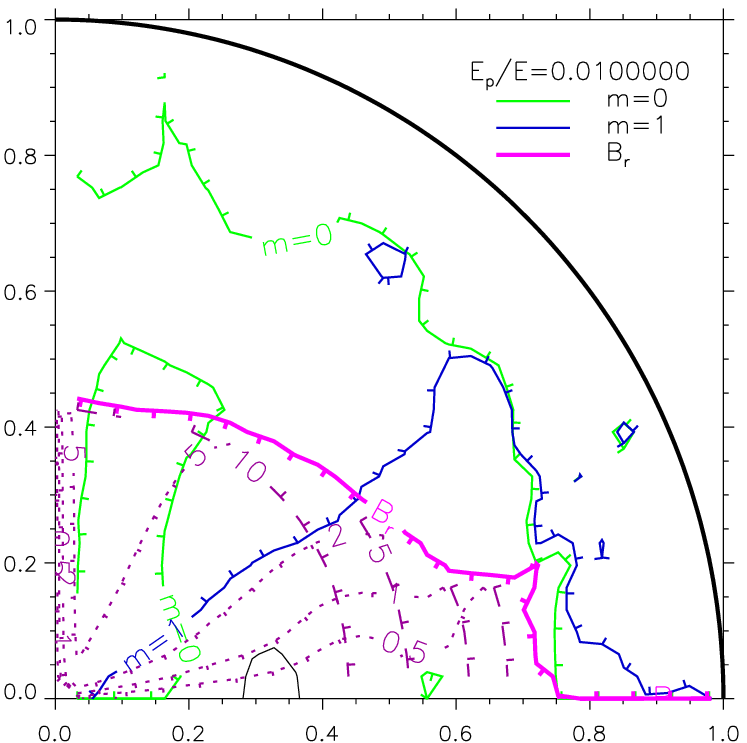}
\includegraphics[width=0.246\hsize,angle=0]{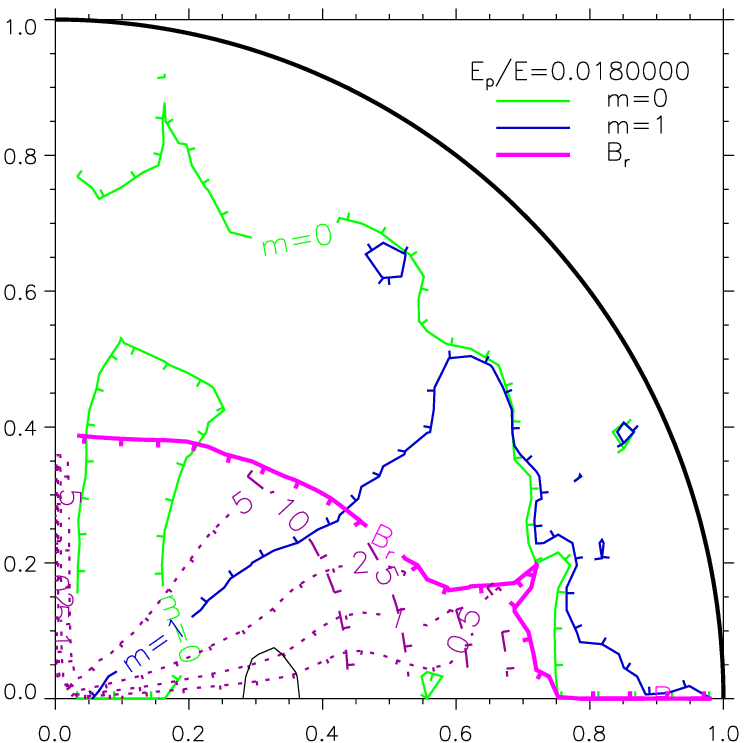}
\includegraphics[width=0.246\hsize,angle=0]{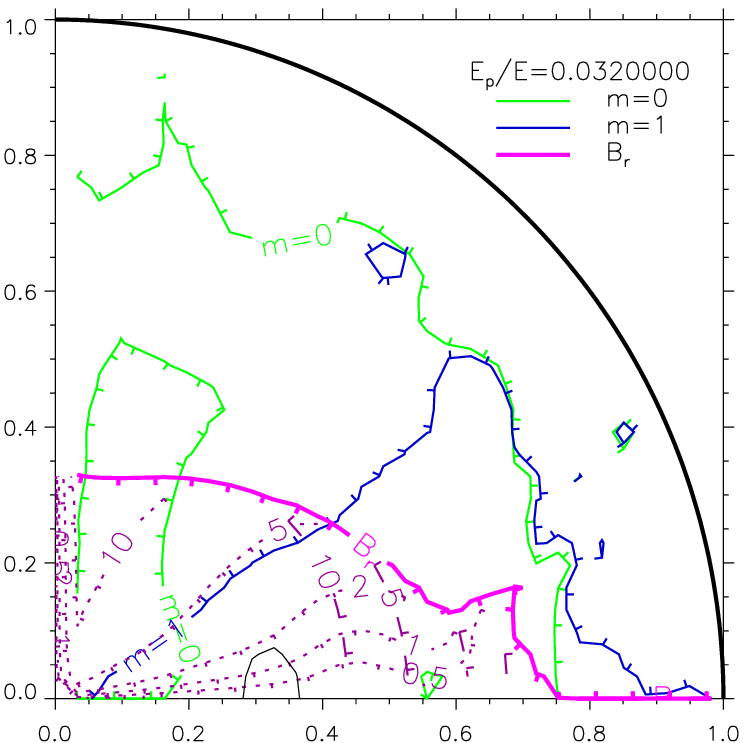}
\includegraphics[width=0.246\hsize,angle=0]{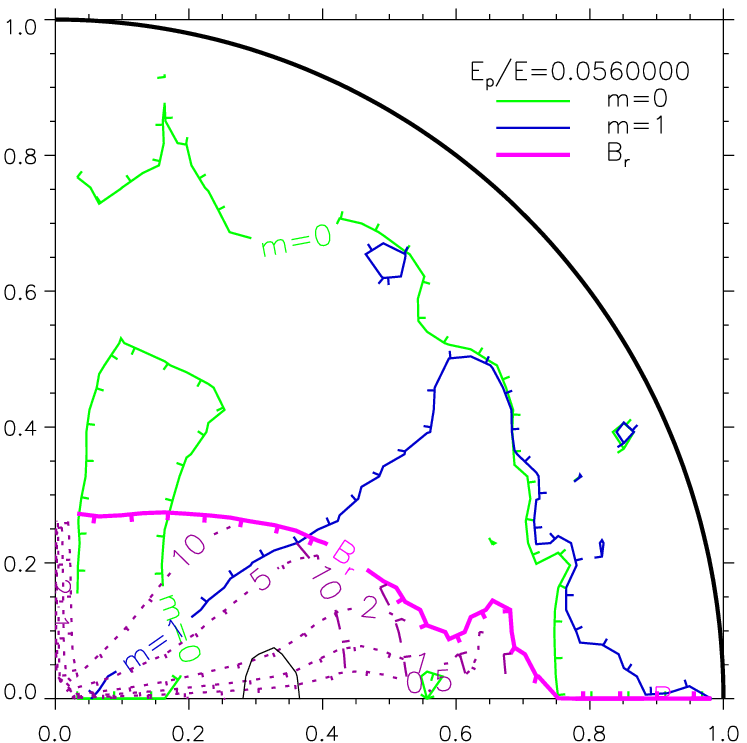}
\includegraphics[width=0.246\hsize,angle=0]{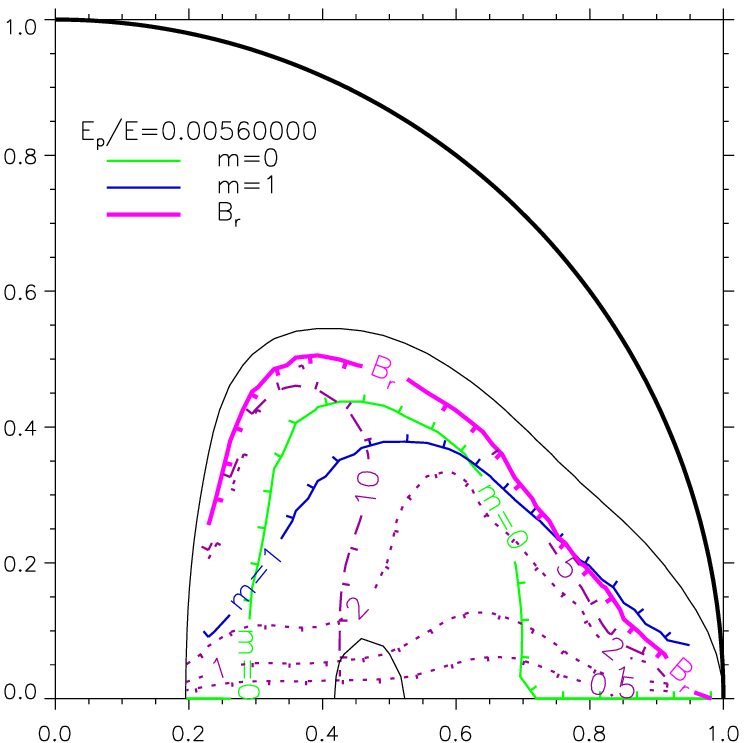}
\includegraphics[width=0.246\hsize,angle=0]{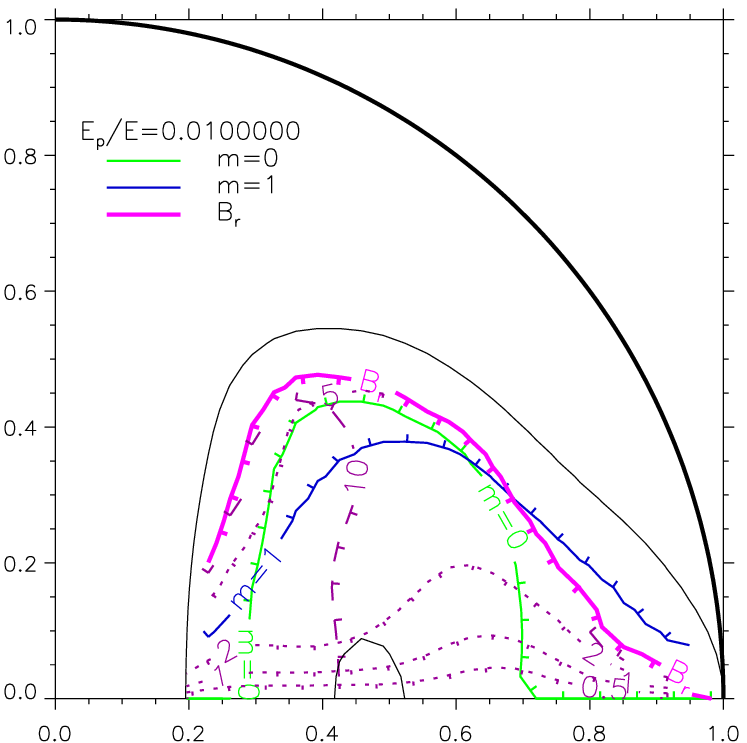}
\includegraphics[width=0.246\hsize,angle=0]{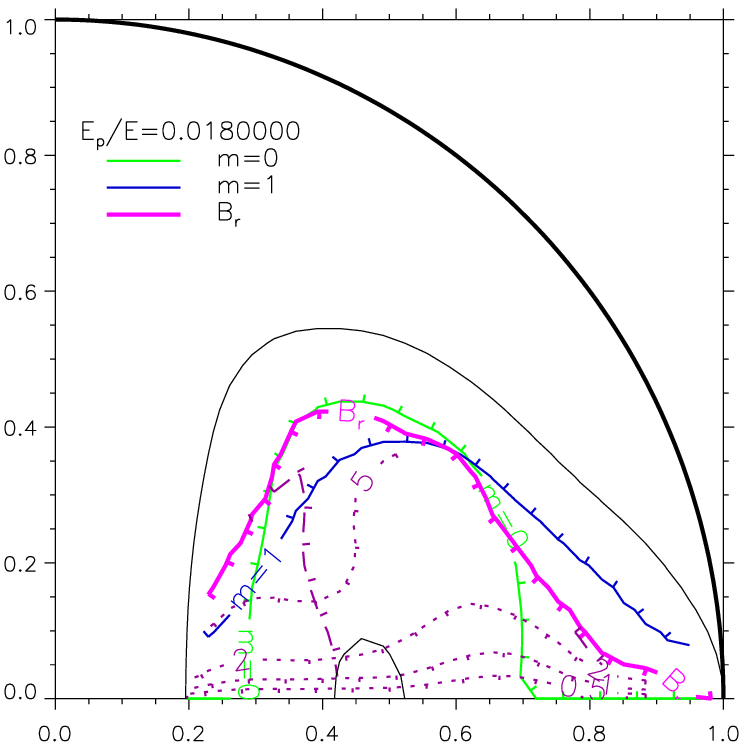}
\includegraphics[width=0.246\hsize,angle=0]{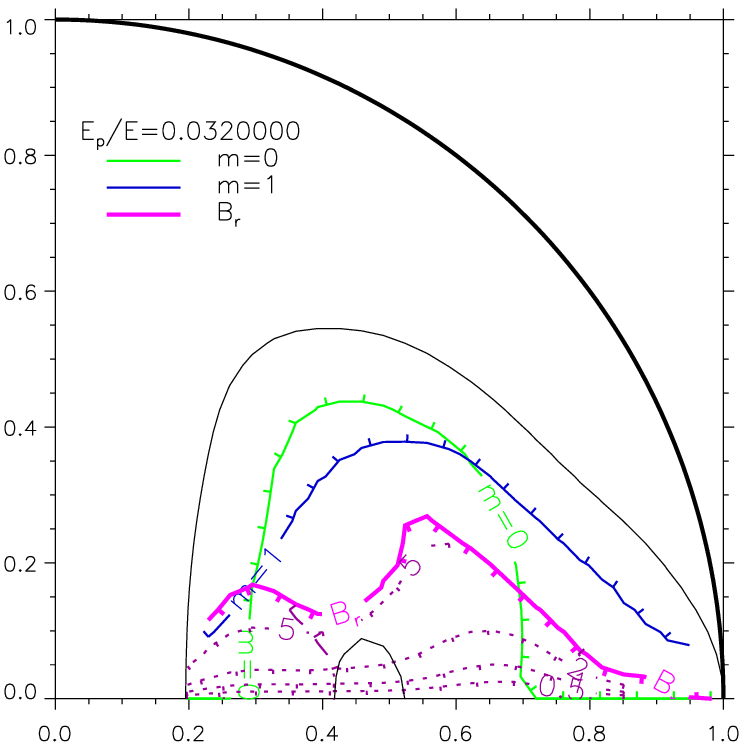}
\includegraphics[width=0.246\hsize,angle=0]{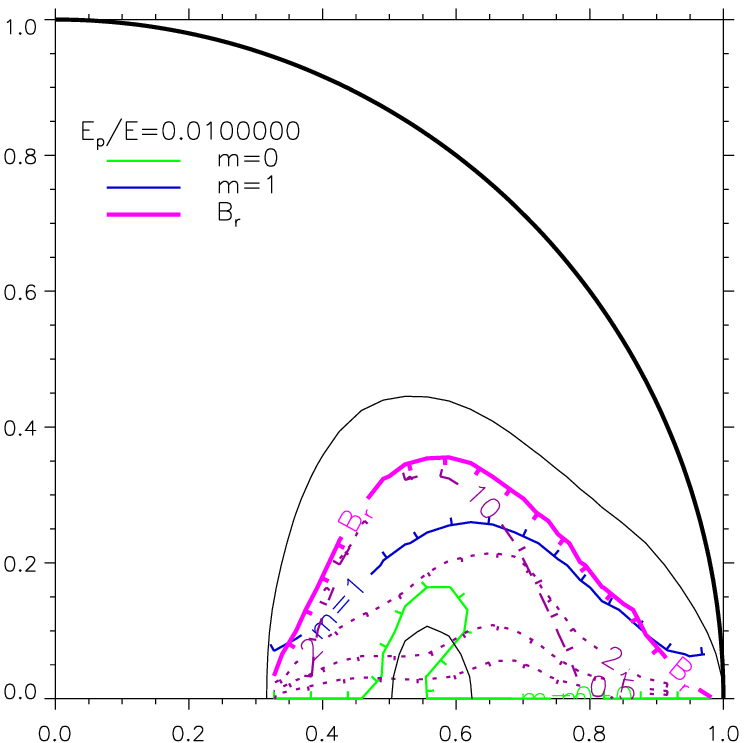}
\includegraphics[width=0.246\hsize,angle=0]{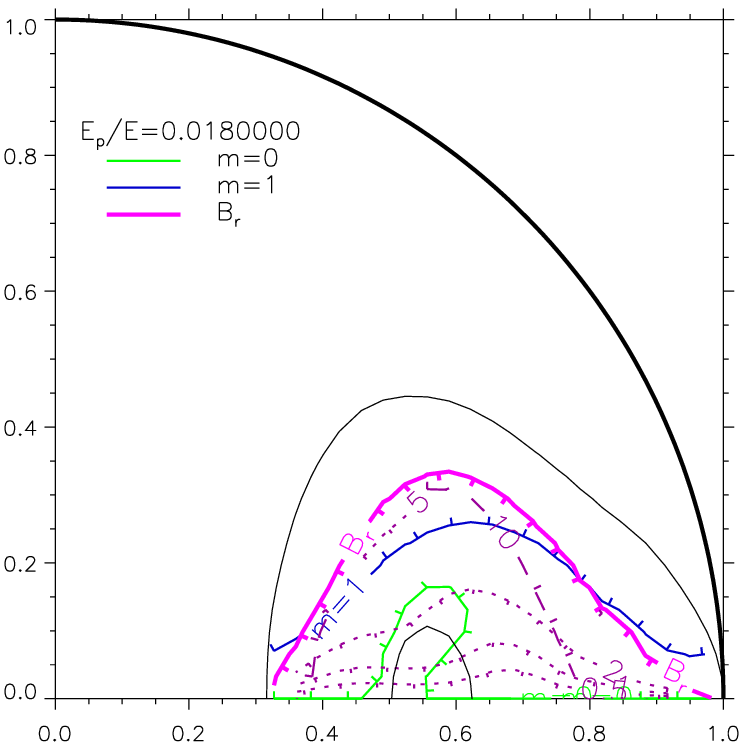}
\includegraphics[width=0.246\hsize,angle=0]{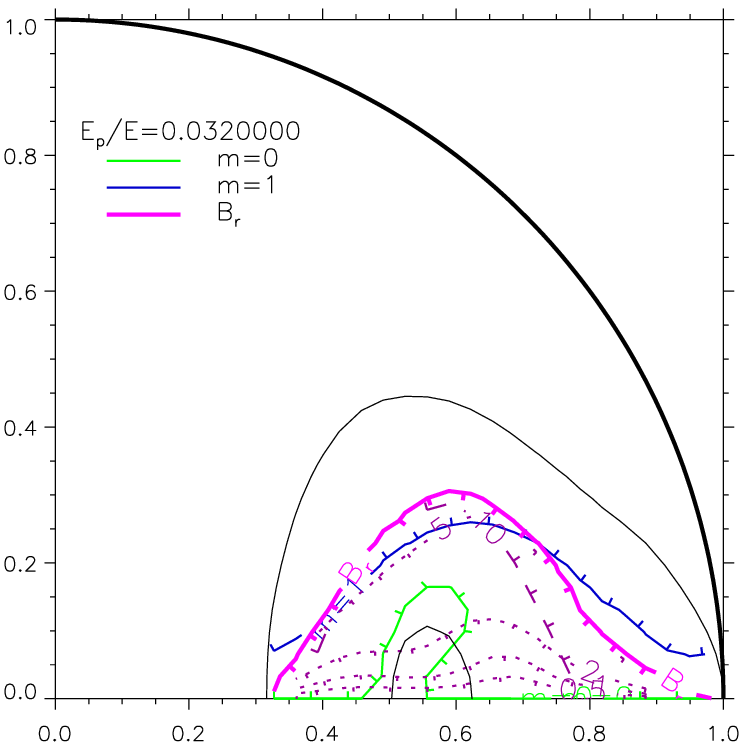}
\includegraphics[width=0.246\hsize,angle=0]{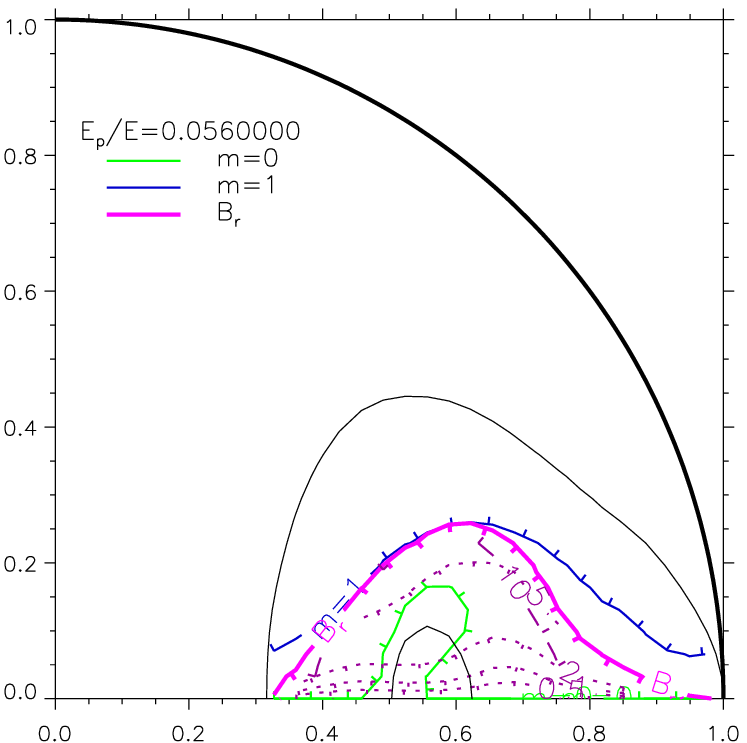}
\caption{Stability lines for various values of $E_{\rm p}/E$ at three points in the fiducual simulation with $E/U=1/400$. Top row: $r_{\rm n}=0.33R$ and $E_{\rm p}/E = 0.01$, $0.018$,$0.032$ and $0.056$; middle row: $r_{\rm n}=0.47R$ and $E_{\rm p}/E = 0.0056$, $0.01$, $0.018$, and $0.032$; bottom row: $r_{\rm n}=0.58R$ and $E_{\rm p}/E = 0.01$, $0.018$, $0.032$ and $0.056$. See fig.~\ref{fig:stabcond} for an explanation of the various lines plotted.}
\label{fig:gall.fracs}
\end{figure*}

These lower limits on $E_p/E$ are, as discussed in section \ref{sec:torwithaxial}, dependent on the absolute field strength which I parametrize as $E/U$, the ratio of magnetic to thermal energies. Changing the magnetic field energy in the fiducial simulation by a factor of $10$ so that $E/U = 4000$ and repeating the exercise above results in stability at a value of $E_p/E$ a factor of $10$ lower, as can be seen in fig.~\ref{fig:weak_field}, where the ratios $E_p/E$ and $E/U$ are both a factor of $10$ lower than in the second plot of the second row of fig.~\ref{fig:gall.fracs}, but the result is near-identical stability lines. The only significant difference between the two is that the unstable wavelengths are a factor $\sqrt{10}$ lower in the weaker field case.

\begin{figure}
\includegraphics[width=1.0\hsize,angle=0]{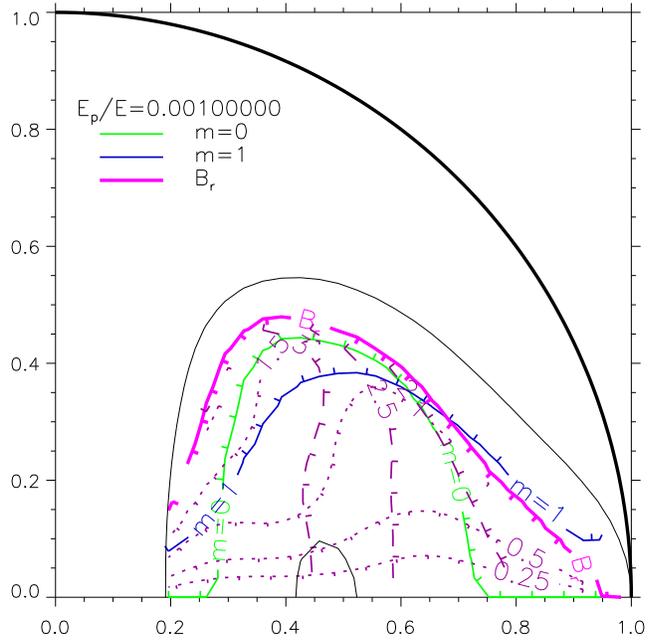}
\caption{Stability lines for $r_{\rm n}=0.47R$ and $E_{\rm p}/E=0.001$ in the weak-field case ($E/U=1/4000$). The resemblance with the $E_p/E=0.01$, $E/U=1/400$ plot in fig.~\ref{fig:gall.fracs} is obvious, the only significant difference is that the unstable wavelengths are a factor $\sqrt{10}$ shorter in the weaker field case. See fig.~\ref{fig:stabcond} for an explanation of the various lines plotted.}
\label{fig:weak_field}
\end{figure}

\subsubsection{Simulations with different $E_{\rm p}/E$}\label{sec:torsims}

To verify this assessment above, it is possible to use these altered $E_{\rm p}/E$ configurations as the initial conditions for simulations. I use here simulations with $E/U=1/400$. First, the star is allowed to relax to equilibrium by keeping the magnetic field constant and letting the pressure field adjust until it balances the Lorentz force. After ten or so sound crossing times, this relaxation has taken place. A small white-noise perturbation is then added to the density field, and the magnetic field is allowed to change, i.e. the induction equation is switched on (this defines $t=0$ in the following plots). In fig.~\ref{fig:mmodes-tor-g5}, the amplitudes of the azimuthal modes $m=0$ to $4$ (an r.m.s. integration over the meridional plane of velocity component $v_\theta$) are plotted against time for simulations with the following values of $E_{\rm p}/E$: $0.01$, $0.018$, $0.032$ and $0.056$, the initial conditions having been taken from the fiducial run at $r_{\rm n}/R=0.33$ before having the poloidal energy ratios changed by hand. It can be seen in the figure that the threshold is at $E_{\rm p}/E \approx 0.032$.

\begin{figure}
\includegraphics[width=1.0\hsize,angle=0]{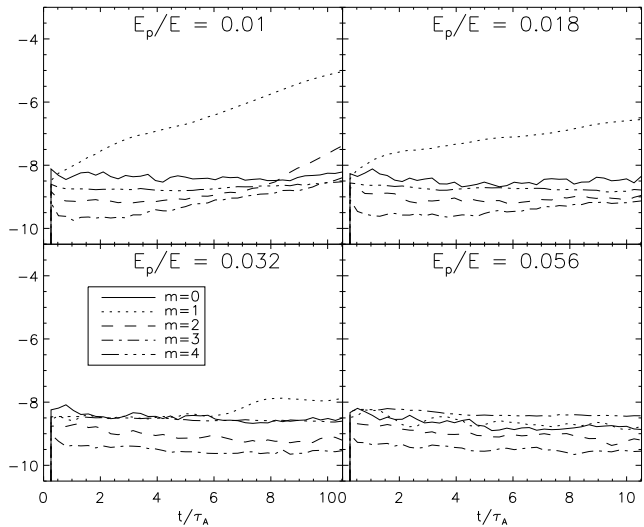}
\caption{Log amplitudes of modes $0 \le m \le 4$ (solid, dotted, dashed, dot-dashed and dot-dot-dot-dashed respectively) in simulations with $r_{\rm n}/R=0.33$ and $E_{\rm p}/E = 0.01$, $0.018$, $0.032$ and $0.056$. The stability threshold appears to be around $0.032$, below which the $m=1$ mode is the dominant mode.}
\label{fig:mmodes-tor-g5}
\end{figure}
\begin{figure}
\includegraphics[width=1.0\hsize,angle=0]{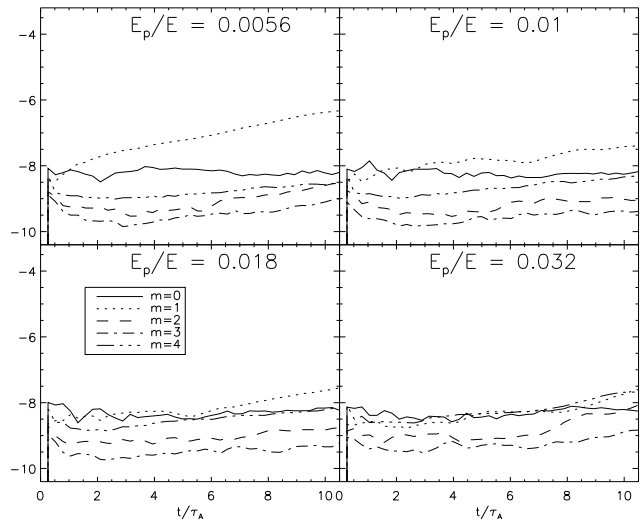}
\caption{Log amplitudes of modes $0 \le m \le 4$ in simulations with $r_{\rm n}/R=0.47$ and $E_{\rm p}/E = 0.0056$, $0.01$, $0.018$ and $0.032$. The stability threshold appears to be around $0.01$ or $0.018$.}
\label{fig:mmodes-tor-g65}
\end{figure}
\begin{figure}
\includegraphics[width=1.0\hsize,angle=0]{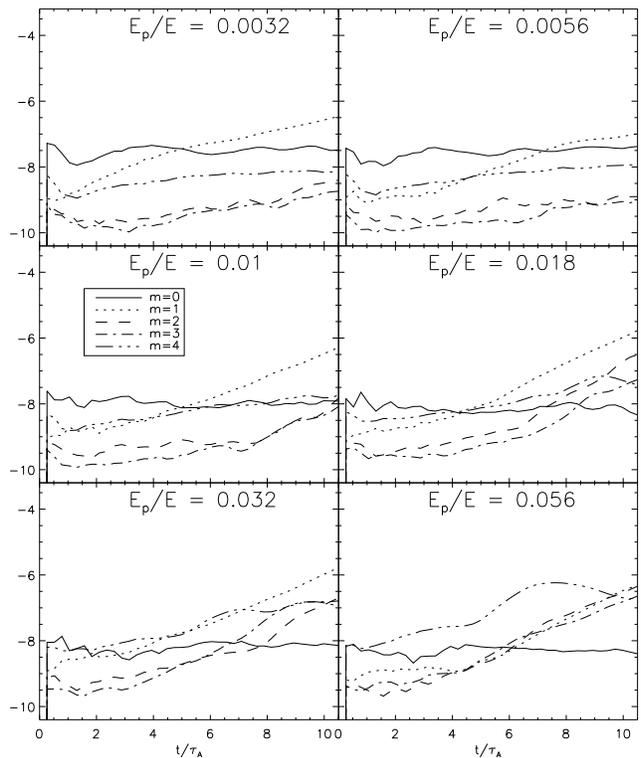}
\caption{Log amplitudes of modes $0 \le m \le 4$ in simulations with $r_{\rm n}/R=0.58$ and $E_{\rm p}/E = 0.0032$, $0.0056$, $0.01$, $0.018$, $0.032$, and $0.056$. The $m=0$ mode appears to be stable at all ratios considered here, whilst the $m=1$ mode appears to stabilise at around $E_{\rm p}/E = 0.056$ (see text).}
\label{fig:mmodes-tor-g90}
\end{figure}

This exercise was then repeated with the equilibria of different values of $r_{\rm n}/R$: $0.47$ and $0.58$: see figs.~\ref{fig:mmodes-tor-g65} and \ref{fig:mmodes-tor-g90}. The former is easy enough to interpret: the threshold is around $E_{\rm p}/E = 0.01$. The small contamination we see in $m=4$ from the geometry of the computational box is visible here but much greater in the $r_{\rm n}/R=0.58$ simulations, presumably because at higher $r_{\rm n}$ the field has a stronger interaction with the sides of the computational box. Here, the ratio $0.056$ simulation, which we know to be stable or almost so because the fiducial run has a ratio $0.066$, has significant $m=4$ contamination but we can see that the $m=1$ mode is stable at least for the first few Alfv\'en crossing times, so it looks like the critical ratio is somewhere between $0.032$ and $0.056$. Surprisingly, the $m=0$ mode seems stable at all ratios presented in the figure. This is at odds to the conclusions of section~\ref{sec:tortayler}, where we found that whilst $m=1$ is stable above $0.01$, the $m=0$ mode should be unstable up to $E_{\rm p}/E=0.056$ or thereabouts. The reason for this discrepancy is not immediately clear, but may have something to do with effects ignored in this study such as curvature effects - the unstable region at this value of $r_{\rm n}$ is near the equator where these could be important.

At this juncture it is necessary to look at the minimum and maximum wavelengths plotted in fig.~\ref{fig:gall.fracs} to check that the simulations have sufficient resolution, as it is conceivable that a $E_{\rm p}/E$ ratio which is in reality unstable could appear to be stable in a simulation simply because the unstable wavelengths are too low to be resolved. Looking at the top row, where the $E_{\rm p}/E = 0.032$ simulation is the first which appears stable, we see that the unstable region near the axis of symmetry has a maximum unstable wavelength of more than $10$ grid spacings ($10dx\approx0.3R$, about the size of the unstable region) and a minimum wavelength of greater than $5dx$. Now, in previous simulations (e.g. \citealt{Braithwaite:2006a}) it was found that this high-order code can resolve modes of wavelength $8dx$ almost perfectly, so it seems unlikely that a $\lambda\approx5dx$ mode is entirely suppressed. Looking at the middle and bottom rows, where the $E_{\rm p}/E = 0.01$ and $0.032$ simulations respectively are the first which appear stable in the simulations, the minimum wavelengths in the $m=1$ unstable regions are also around $5dx$ in both cases.

Ideally one would now perform simulations with different field strengths, i.e. different values of $E/U$ from the value $1/400$ used above, in order to check the analytic prediction that the critical poloidal energy fraction $(E_{\rm p}/E)_{\rm crit} \propto E/U$. However, we see from fig.~\ref{fig:weak_field} that the unstable wavelengths will be badly resolved or not resolved at all, and therefore that significantly higher numerical resolution would be required. At higher field strengths we are no longer in the physically interesting weak-field regime, as the structure of the star becomes significantly non-spherical. Checking this result numerically at the higher resolution required will be left for the future. It is not inconceivable of course that some other instability not considered here could become relevant at lower field strengths.

\subsection{Stability of a predominantly poloidal field}
\label{sec:poldom}


We have seen that an axisymmetric configuration with much stronger a toroidal component than poloidal can be stable. However, we have reason to believe that the opposite is not true; we expect that a field with a much stronger poloidal component will make a transition to non-axisymmetric equilibrium.

We can now apply the stability conditions (\ref{eq:polfrac}) and (\ref{eq:wright}) to the fiducial simulation at the three points in time used in the previous section, i.e. when $r_{\rm n}/R = 0.33$, $0.47$ and $0.58$. We find that the thresholds in $E_{\rm p}/E$ are $0.71$, $0.48$ and $0.69$ using (\ref{eq:polfrac}). How to calculate Wright's (1973) threshold is not obvious; one way is to look at conditions near the neutral line and use $s_0 B_{\rm p}/s$ for $B_{\rm p}(s_0)$ and set $s_0$ to $r_{\rm n}$, and use the value of $B_{\rm t}$ on the neutral line. This gives thresholds of $0.91$, $0.80$ and $0.91$. Wright points out that his stability condition is a necessary condition, and should be interpreted as a lower bound on the toroidal field required for stabilisation.

To test the strength of toroidal field required for stabilisation, simulations analogous to those presented in section~\ref{sec:torsims} were performed, this time with high values of $E_{\rm p}/E$. The pertubation added in this case differs from the white noise used in the previous section: a large-scale pertubation was added to azimuthal modes $m=2$ to $8$. In figs.~\ref{fig:mmodes-pol-g5}, \ref{fig:mmodes-pol-g65} and \ref{fig:mmodes-pol-g90} the amplitudes of modes $m=0$ to $m=4$ are plotted against time for different poloidal energy ratios at different values of $r_{\rm n}$.

The thresholds in all cases seem to be at $E_{\rm p}/E \approx 0.8$. In the $r_{\rm n}=0.33R$ simulations (fig.~\ref{fig:mmodes-pol-g5}), at $0.8$ (and at $0.7$), an oscillatory behaviour can be seen in the $m=2$ mode, indicating stability. At $0.9$, the $m=2$ mode is dominant; at $0.944$ both $m=2$ and $m=3$ grow (the $m=4$ mode is probably just numerical contamination at these ratios, but will become properly unstable above some threshold in $E_{\rm p}/E$).
Looking at fig.~\ref{fig:mmodes-pol-g5}, where $r_{\rm n}=0.47R$, we see that $E_{\rm p}/E = 0.7$ is clearly stable; at $0.8$ the $m=2$ appears stable and then unstable -- this is probably because the magnetic field is just below the threshold and is taken over it by continuing secular evolution, which weakens the toroidal component faster than the poloidal. It is visible in the plots that at higher poloidal energy ratios that the higher azimuthal modes become unstable. Similarly we see in fig.~\ref{fig:mmodes-pol-g5} ($r_{\rm n}=0.58R$) that the stability threshold is very close to $0.8$. Very clear here is the stability (hence oscillations) of $m=3$ at $0.9$ but unstable growth at $E_{\rm p}/E=0.944$. Of course, oscillations follow unstable growth when saturation, i.e. a new non-axisymmetric equilibrium, is reached.

\begin{figure}
\includegraphics[width=1.0\hsize,angle=0]{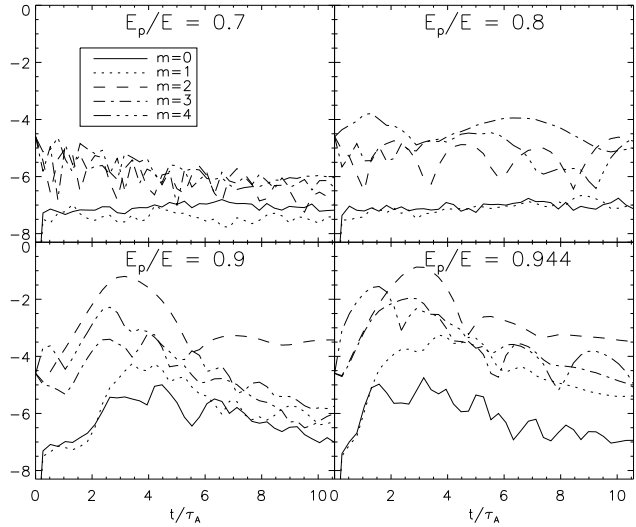}
\caption{Log amplitudes of azimuthal modes $0 \le m \le 4$ (solid, dotted, dashed, dot-dashed and dot-dot-dot-dashed respectively) in simulations with $r_{\rm n}=0.33R$. The stability threshold appears to be between $0.8$ and $0.9$.}
\label{fig:mmodes-pol-g5}
\end{figure}

\begin{figure}
\includegraphics[width=1.0\hsize,angle=0]{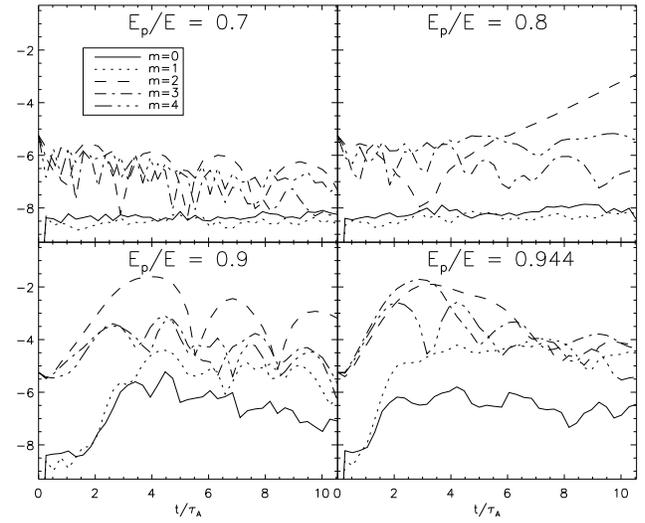}
\caption{Log amplitudes of azimuthal modes $0 \le m \le 4$ in simulations with $r_{\rm n}=0.47R$. The stability boundary is around $0.8$.}
\label{fig:mmodes-pol-g65}
\end{figure}

\begin{figure}
\includegraphics[width=1.0\hsize,angle=0]{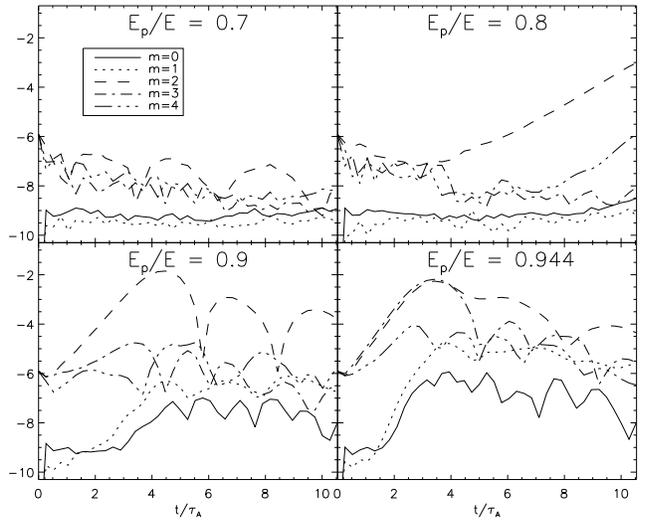}
\caption{Log amplitudes of azimuthal modes $0 \le m \le 4$ in simulations with $r_{\rm n}=0.58R$.}
\label{fig:mmodes-pol-g90}
\end{figure}

A perhaps more elegant way to present the output from the simulations is to plot a map of the field on the surface of the star. Figs.~\ref{fig:map-07000} and \ref{fig:map-higher} are such maps of $B_r$; all are taken from the $r_{\rm n}/R = 0.47$ simulations. Fig..~\ref{fig:map-07000} follows the evolution in time of three simulations with different $E_{\rm p}/E$ ratios; by the end of the simulations new equilibria have been reached in the unstable cases. These maps confirm that $E_{\rm p}/E = 0.7$ is stable (apart from a small $m=4$ numerical contamination), that $0.8$ is stable at first and then succumbs to the $m=2$ mode, and that the $0.9$ field is unstable mainly to the $m=2$ mode and eventually finds its way into a new equilibrium. Fig.~\ref{fig:map-higher} shows the equilibria reached in simulations with even higher $E_{\rm p}/E$ ratios; evidently, the higher the ratio, the more complex the resulting equilibrium.

\begin{figure*}
\includegraphics[width=0.33\hsize,angle=0]{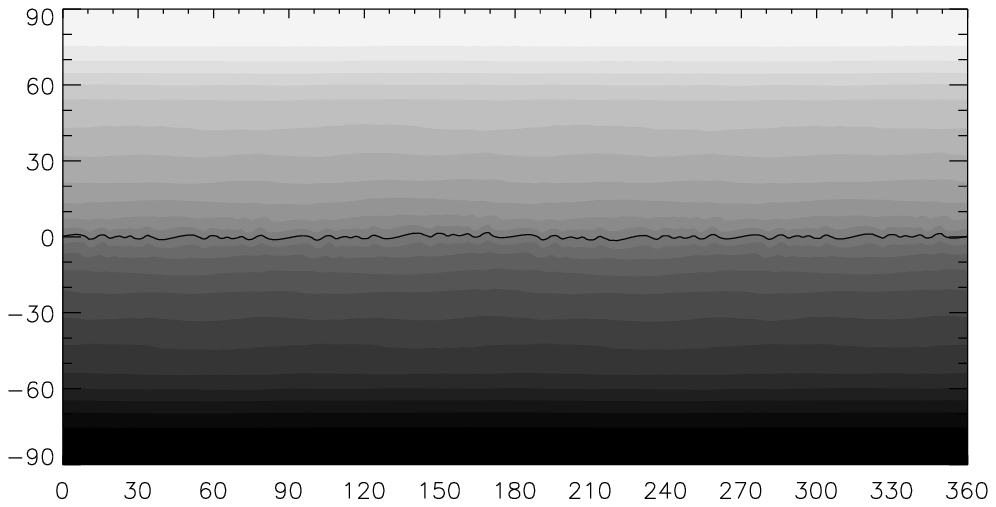}
\includegraphics[width=0.33\hsize,angle=0]{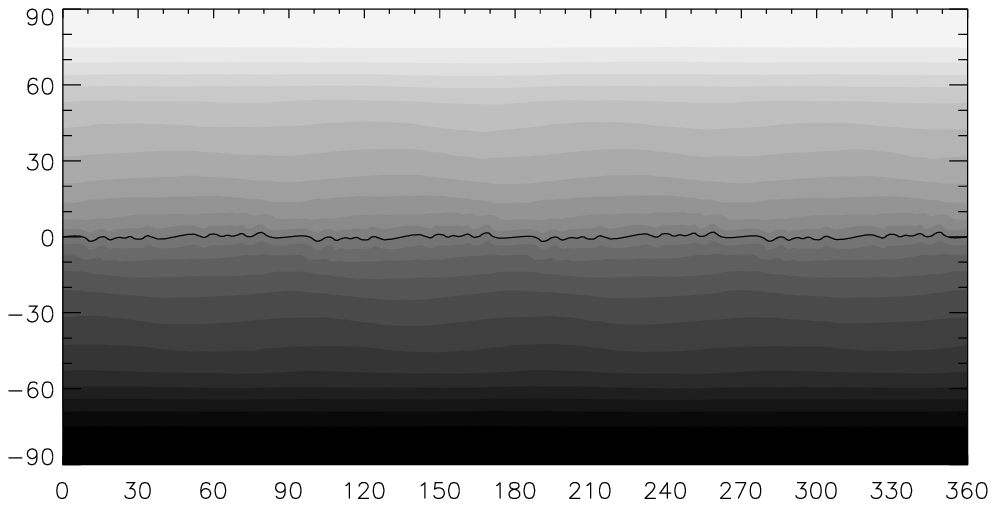}
\includegraphics[width=0.33\hsize,angle=0]{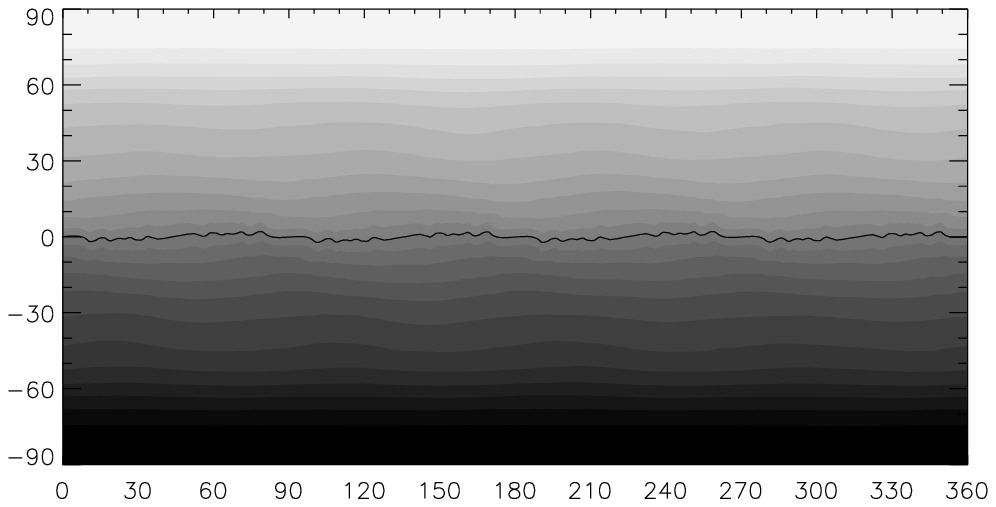}
\includegraphics[width=0.33\hsize,angle=0]{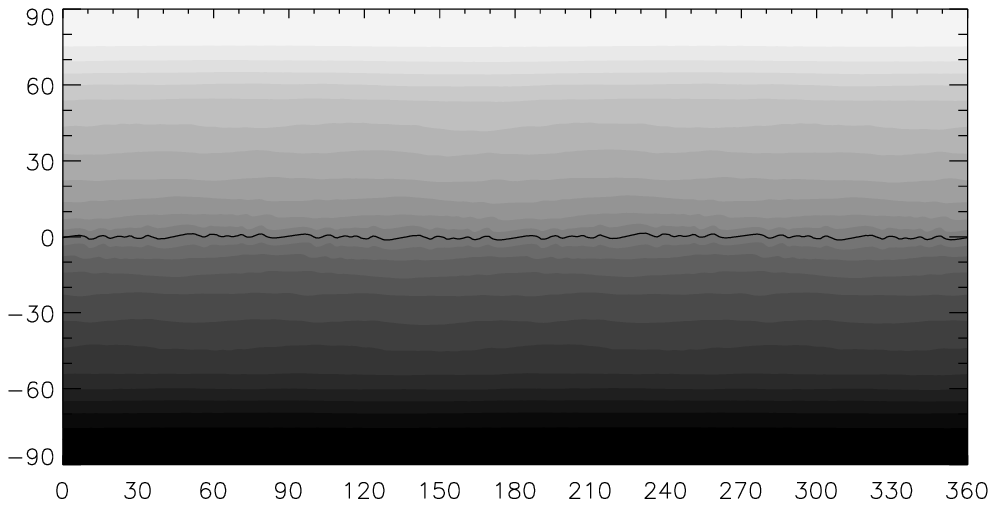}
\includegraphics[width=0.33\hsize,angle=0]{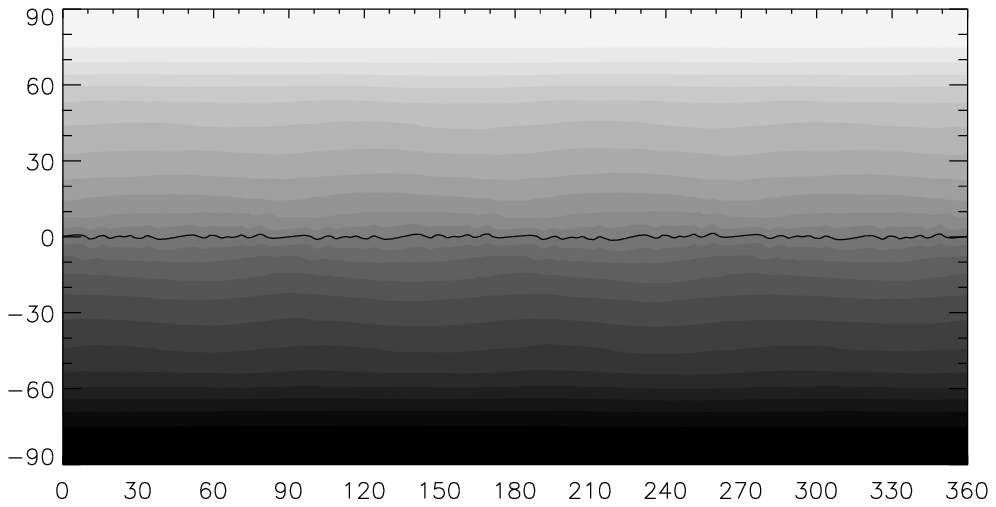}
\includegraphics[width=0.33\hsize,angle=0]{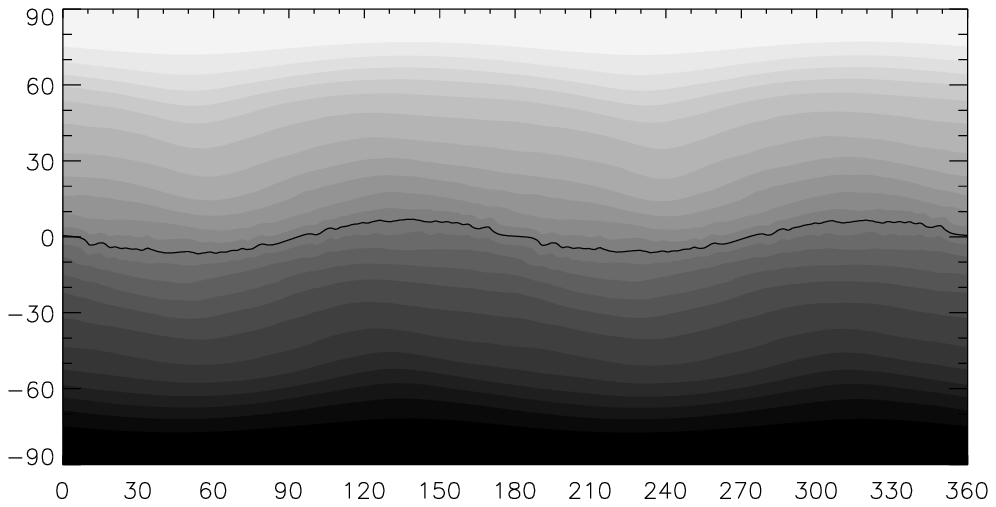}
\includegraphics[width=0.33\hsize,angle=0]{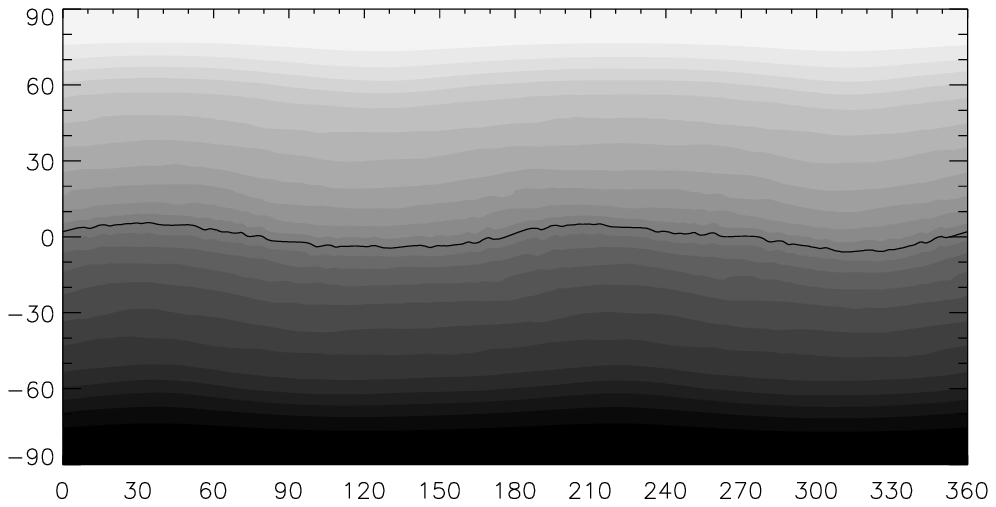}
\includegraphics[width=0.33\hsize,angle=0]{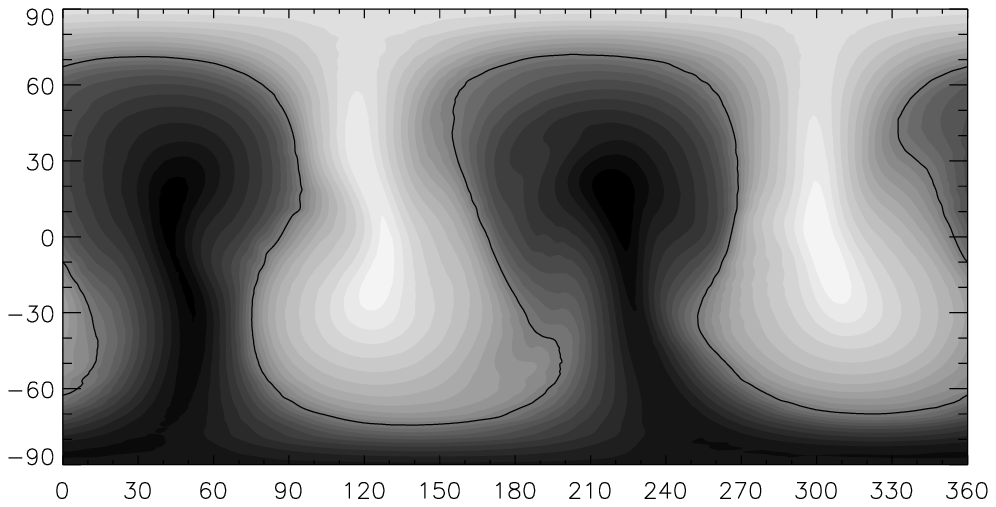}
\includegraphics[width=0.33\hsize,angle=0]{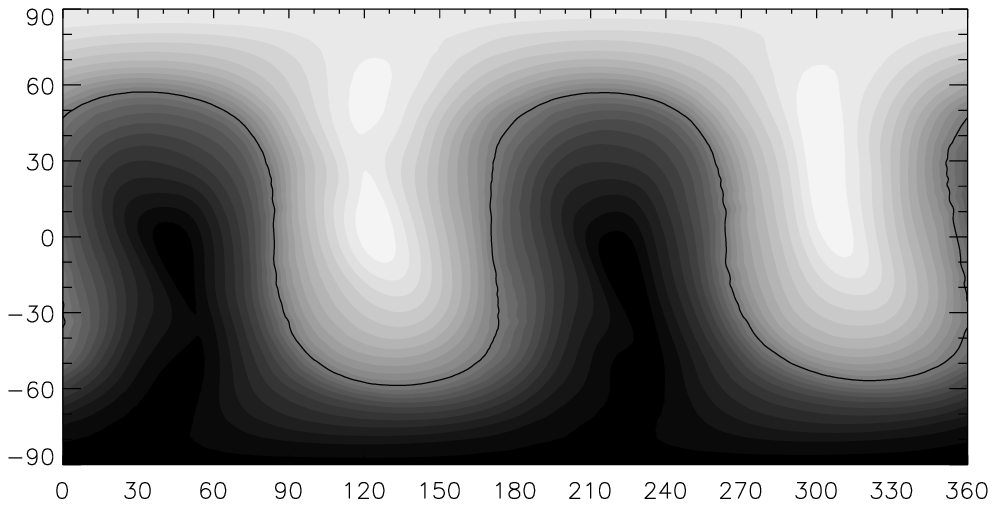}
\caption{{\mk Behaviour of the magnetic field when the poloidal component is stronger than the toroidal, near the maximum poloidal/toroidal ratio.} Maps of the radial magnetic field $B_r$ on the stellar surface at three points in time ($t/\tau_{\rm A} = 2.9$, $5.8$ and $10.5$; plates on the left, middle and right respectively) of the simulations with $r_{\rm n}/R = 0.47$ and $E_{\rm p}/E = 0.7$, $0.8$ and $0.9$ (top, middle and bottom rows respectively). White and black represent the strongest positive and negative $B_r$, and the black line shows the position of $B_r=0$, which is initially at the equator. The simulation on the top row shows stability; on the middle row we see initial stability followed by an $m=2$ mode, and the simulation with $E_{\rm p}/E = 0.9$ shows a strong growth of the $m=2$ mode which then saturates after a few Alfv\'en crossing times as the field settles into a new equilibrium.}
\label{fig:map-07000}
\end{figure*}

\begin{figure*}
\includegraphics[width=0.33\hsize,angle=0]{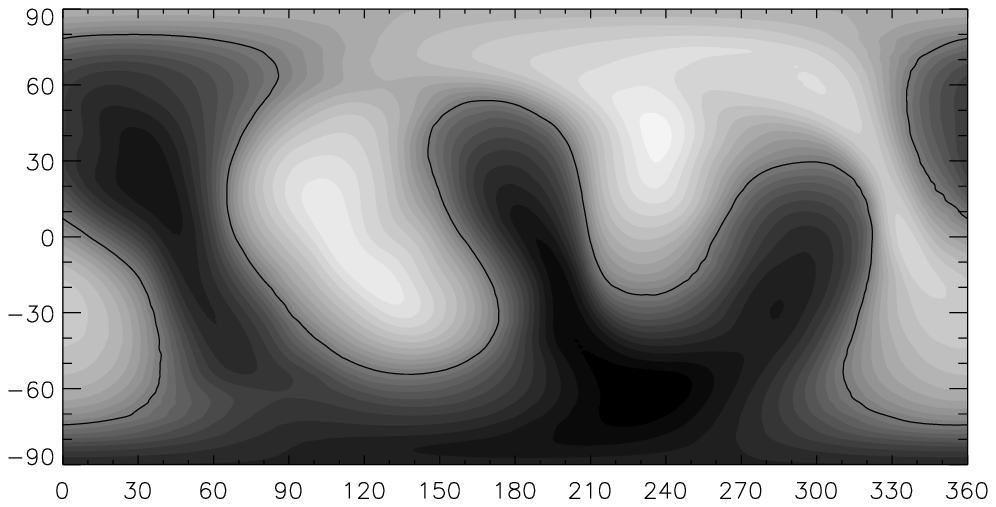}
\includegraphics[width=0.33\hsize,angle=0]{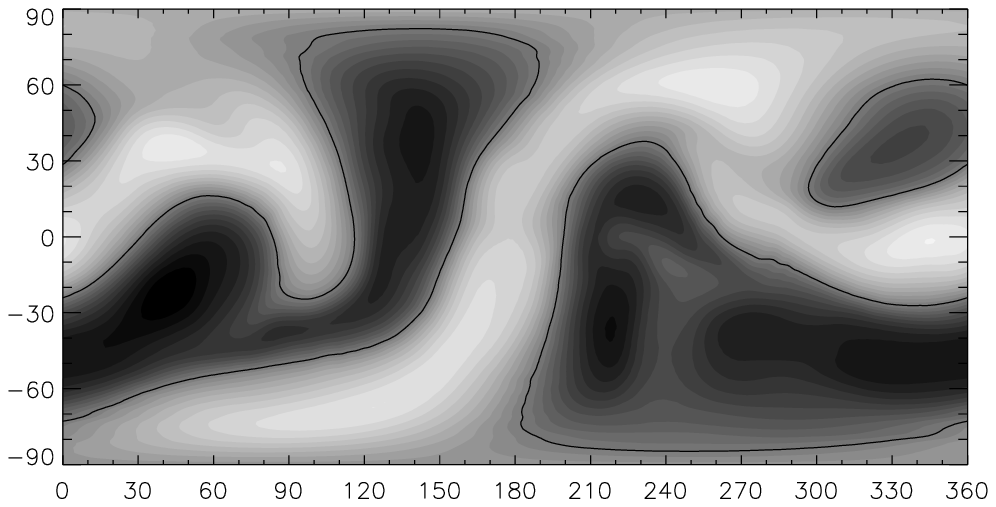}
\includegraphics[width=0.33\hsize,angle=0]{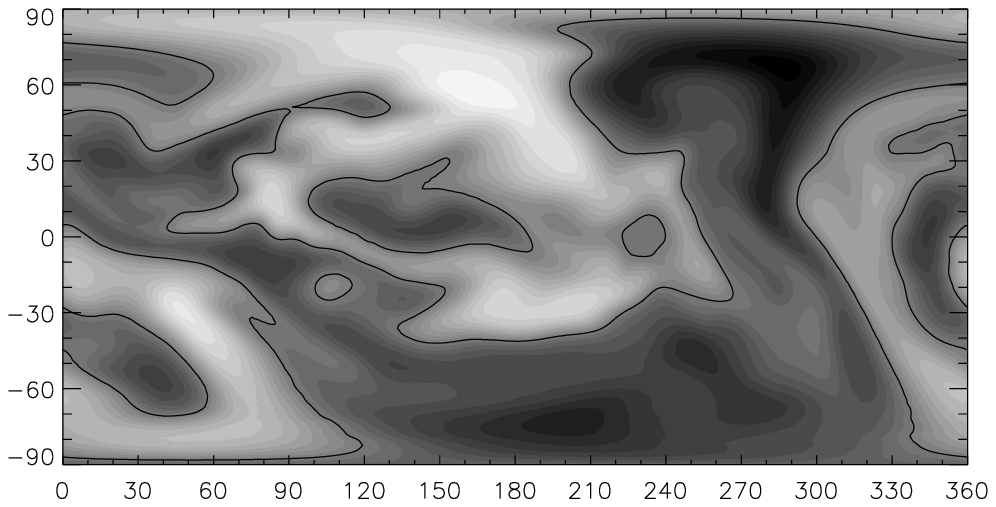}
\caption{As fig.~\ref{fig:map-07000} but all three $B_r$ maps are at one point in time ($t/\tau_{\rm A} = 10.5$) in three different simulations with $E_{\rm p}/E = 0.968$, $0.99$ and $1.0$. Evidently, at higher initial $E_{\rm p}/E$ ratios, the non-axisymmetric equilibria reached are more complex.}
\label{fig:map-higher}
\end{figure*}

It seems then that a magnetic field becomes unstable as the poloidal energy fraction exceeds about $80\%$, and that this threshold depends only weakly on the central concentration of the field (i.e. on $r_{\rm n}$). Just above this threshold, the only unstable mode is $m=2$, which is reassuringly well resolved by simulations at this resolution. At higher values of $E_{\rm p}/E$ higher modes become unstable, which is exactly what we expect because higher azimuthal modes have to do more work against the toroidal field, in proportion to the value of $m$. Therefore, the $m=3$ mode becomes unstable with a toroidal field $2/3$ the strength of that at the $m=2$ stability threshold, so that if the $m=2$ threshold is $E_{\rm p}/E\approx 0.8$, then the $m=3$ threshold will be at $E_{\rm p}/E \approx 1-(1-0.8)(2/3)^2 = 0.911$ (ignoring the small change in poloidal field strength resulting from the change in $E_{\rm_p}/E$ at constant $E$). This can be seen for instance in fig.~\ref{fig:mmodes-pol-g90} and confirms the $m$ dependence of in (\ref{eq:polfrac}).

The stability thresholds found from these simulations agree only qualitatively with the predictions (\ref{eq:polfrac}) and (\ref{eq:wright}). This probably has to do with effects not included in the analysis, such as the the toroidal geometry of the flux tube, the fact that the poloidal field lines are not circular in the neighbourhood of the neutral line - in fact they are decidedly elliptical, and that the toroidal field falls off away from the neutral line.

It is informative to think about this evolution into a new equilibrium in terms of the magnetic helicity, defined as $\int {\mathbf B}\cdot{\mathbf A} dV$ where ${\mathbf A}$ is the vector potential. As the field evolves on a dynamical timescale, helicity is conserved and the field can be thought of as evolving into the lowest energy state for that value of helicity. Essentially, helicity can be thought of as the product of toroidal and poloidal fluxes, so that as we go to higher $E_{\rm p}/E$ ratios, the helicity falls, and below some threshold the lowest energy state is non-axisymmetric. The initial equilibrium is essentially a twisted flux tube lying in a circle around the equator of the star, and the transition to non-axisymmetric equilibrium is a matter simply of stretching this flux tube into a more complex arrangement. In the process of stretching the tube, the toroidal component (i.e. the component parallel to the axis of the tube, the neutral line) is amplified  (since the tube becomes narrower), and the poloidal component becomes weaker, eventually bringing the two components to roughly equal strengths, because the energy minimum for a given helicity, i.e. for a given product of toroidal and poloidal field strengths, will have the two components roughly equal to each other. At higher $E_{\rm p}/E$ ratios therefore, more stretching is required to make the two components equal. This can be clearly seen in fig.~\ref{fig:map-higher}. Now, in the case where $E_{\rm p}/E=1$ the field has zero magnetic helicity and no amount of flux-tube stretching can result in an equilibrium. However, there are diffusive processes at work which can either create helicity or split the one original flux tube into two or more tubes which can have helicity of different signs and which add up to zero, although it is likely that a lot of time will pass before any equilibrium is reached and the energy of the equilibrium will be very much lower than the original energy.

\section{Conclusions and discussion}
\label{sec:discussion}

In this paper, I have looked at the lower and upper limits on the fractions of energy in the poloidal and toroidal components of an axisymmetric magnetic field. To find these limits, I took the output from a simulation where a `turbulent' initial magnetic field evolves into an axisymmetric equilibrium, changed the relative strengths of the poloidal and toroidal components by hand, and used that as the initial conditions for new simulations. This was supplemented with more analytic methods including the necessary and sufficient stability conditions found by \citet{Tayler:1973} (it is incidentally found that four of his six conditions are always met at every point in the star). The two methods are in broad agreement.

The result of this investigation is that while the upper limit on the poloidal energy fraction $E_{\rm p}/E$ is around $80\%$, the lower limit depends on factors such as the radius of the neutral line $r_{\rm n}$ and can be between $1\%$ and roughly $5\%$ for a star constructed from a polytrope of index $n=3$ (which approximates to an upper-main-sequence star) and where the ratio of magnetic to thermal energies $E/U=1/400$ . This lower limit is expected to be proportional to the ratio $E/U$, so that $(E_{\rm p}/E)_{\rm crit} \sim 10 E/U$. These limits will also depend on other factors not explicitly explored here, such as the equation of state and density profile of the star, but these should not affect the results in more than a modest quantitative manner; however in a NS we might expect a lower limit of $(E_{\rm p}/E)_{\rm crit} \sim 10^3 E/U$. The upper limit found here broadly confirms what was expected from the analysis in Paper II and from the analyses of \citet{Wright:1973} and \citet{MarandTay:1974}, who found that the toroidal field must be at least about a quarter of the strength of the poloidal field. The lower limit on $E_{\rm p}/E$ had not been examined previously.

The question of what ratios we actually expect to find in nature has not yet been answered. It will depend on the state of the magnetic field left over from the convective protostellar phase (whether we are looking at main-sequence stars, white dwarfs or neutron stars) and on the subsequent secular evolution. In the fiducial simulation described, $E_{\rm p}/E$ does fall to around $0.046$ and is below $0.1$ for most of the period of diffusive evolution. Given that a star is generally strongly differentially rotating when it is formed, and that any seed field will be wound up and predominantly toroidal, it seems plausible that the eventual equilibrium could have rather low $E_{\rm p}/E$. A proper study of the effect of initial conditions on the resulting equilibrium will be left for the future.

It remains to looks at some of the implications of these results. The magnetic field of a star can have various effects on its appearance and behaviour. In upper main-sequence stars and white dwarfs, we can directly observe the field on the surface via the Zeeman effect, and it turns out that many, if not most, do have roughly axisymmetric configurations. In neutron stars, we measure the spindown and infer from that the dipole component on the surface. Below the surface of the star, there could be a deeply buried field (with low $r_{\rm n}$ and very low $\Phi_{\rm surf}/\Phi_{\rm p}$) and/or a very strong toroidal component, both of which could in effect `hide' magnetic energy from view. This is of obvious interest in the study of magnetars, neutron stars with observed dipole fields of $10^{14-15}$ gauss whose emission in X rays and $\gamma$ rays is powered by the decay of the magnetic field. The possibility that these stars could contain a large quantity of energy hidden from view could explain the large energy output of these objects. A field of $10^{15}$G contains around $2\,10^{47}$erg in magnetic energy, and flare has been observed which emitted a tenth of that quantity in less than a second. According to the standard flare model, some slow evolution of the magnetic field in the core results in stress build-up in the crust, which eventually results in the crust cracking and a release of energy, but it is difficult to imagine this mechanism releasing a large fraction of the magnetic energy during any one event. A field which is stronger in the core of the star than the poloidal component we see on the surface is one solution to this problem. Another possible solution is that the field is very non-axisymmetric, and that the dipole measured is an underestimate of the average field strength on the surface. Any of these are also potential solutions to the phenomenon of neutron stars with very different observational properties but which occupy the same region on the $P-\dot{P}$ diagram and have therefore the same dipole field strength.

Another effect that the magnetic field has is to deform the star's mass distribution. In the light of results presented above, it seems likely that a star has a predominantly toroidal field, deforming the star into a prolate shape. This causes the star to undergo torque-free precession and the damping of this precession will cause the star's magnetic axis to tend towards orthogonality with the rotation axis. In neutron stars this could potentially happen fairly quickly, faster than the orientation of the axes can be changed by spindown torque, which acts on the spindown timescale. In the millisecond magnetar models, the proto-NS rotates at some significant fraction of break up and this results in a powerful dynamo which creates a strong magnetic field during the first hundred seconds. Since it seems very likely that the equilibrium formed after this dynamo switches off is predominantly toroidal, the star should flip over and emit gravitational waves observable with the next generation of detectors at least as far away as the Virgo cluster. However, the spindown timescale of a NS rotating with $P=1$ms and $B=10^{15}$G is only an hour, so the flipping-over mechanism has to work fairly fast. Since the free precession period of such a star would be $P/\epsilon = 1000$s, where $\epsilon\sim10^{-6}$ is the ellipticity induced by the magnetic field of this magnitude, the damping of the precession would have to be very efficient indeed unless the magnetic equilibrium could be produced already at an angle to the rotation axis. More work is required to address these issues.

{\it Acknowledgements.} The author would like to thank \AA ke Nordlund, Chris Thompson and Henk Spruit for assistance and useful discussions and also Anna Watts for pointing out the importance of this problem.












\bsp

\end{document}